\documentstyle[preprint,aps,eqsecnum,epsfig]{revtex} 
\tighten 
\begin{document} 
\preprint{PITT-98-??; LPTHE-98/54} 
\draft  
\title{\bf DYNAMICAL RENORMALIZATION GROUP RESUMMATION OF FINITE
TEMPERATURE INFRARED DIVERGENCES}   
\author{\bf D. Boyanovsky$^{(a,b)}$, H. J. de Vega$^{(b,a)}$,
R. Holman$^{(c)}$, M. Simionato$^{(b,d)}$} 
\address
{(a) Department of Physics and Astronomy, University of 
Pittsburgh, Pittsburgh  PA. 15260, U.S.A \\ 
(b) LPTHE, Universit\'e Pierre et Marie Curie (Paris VI) et Denis Diderot 
(Paris VII), Tour 16, 1er. \'etage, 4, Place Jussieu, 75252 Paris, Cedex 05, 
France\\
(c) Department of Physics, Carnegie-Mellon University, Pittsburgh, 
PA. 15213, U.S.A. \\
(d) INFN, Gruppo Collegato di Parma, Italy} 
\maketitle 
\begin{abstract} 
We introduce the method of dynamical renormalization group to study
 relaxation and damping out of equilibrium directly in real time and
 applied it to the study of infrared divergences in scalar QED. 
 This method allows a consistent resummation of infrared effects
 associated with the exchange of quasistatic transverse photons and
 leads to anomalous logarithmic relaxation of the form $ e^{-\alpha \,
 T \, t \, \ln[t/t_0]}$  for hard momentum charged excitations. 
This  is in contrast with the usual quasiparticle 
interpretation of charged collective
 excitations at finite temperature 
in the sense of exponential relaxation of a narrow width resonance for
which the width is the imaginary part of the self-energy on-shell. 
In the case of narrow resonances away from thresholds, this approach leads to the usual 
exponential relaxation.
The hard thermal loop resummation
program is incorporated consistently into the dynamical
renormalization group  yielding  a picture of relaxation and damping
phenomena in a plasma in real time that trascends the 
conceptual limitations of the quasiparticle picture and other type of
 resummation schemes.  

\end{abstract} 
\section{Introduction} 
The possibility of studying experimentally the formation and evolution
of the quark gluon plasma at RHIC and LHC motivates 
a deeper understanding of collective excitations in ultrarelativistic
plasmas (for reviews see\cite{qgp}-\cite{lebellac}).  
An important concept in the assessment of whether the quark gluon
plasma achieves local thermodynamic equilibrium  is 
that of the relaxation time scale or damping rate which determines the
lifetime of excitations in the
plasma\cite{kapusta,lebellac,lifetime,htlnonabel}. The concept and
definition of the damping rate of a 
 collective excitation is
associated with a quasiparticle description of these excitations in
the plasma and imply exponential relaxation. The validity of the 
quasiparticle concept requires that the lifetime must be large compared
to the oscillation period of the particular excitation mode. In this
quasiparticle picture the collective excitations 
are described as narrow resonances, their spectral function is of the
Breit-Wigner form and the damping rate is obtained from the width of
this resonance. For weakly interacting 
quasiparticles the narrow resonance (quasiparticle) approximation is
expected to be reliable and the damping rate or 
lifetime is obtained from the imaginary part of the self-energy on the
mass shell of the collective
excitation\cite{kapusta,lebellac,lifetime,htlnonabel}.  

Early attempts to calculate the damping rates of quasiparticles in
lowest order perturbation theory obtained gauge 
dependent and unphysical results\cite{rob1}. Braaten and Pisarski\cite{rob2}-\cite{blaizot1} introduced a 
resummation scheme (the resummation of
 the hard thermal loops or HTL) that incorporates the screening corrections in a gauge invariant manner and render finite
transport cross sections\cite{transport}. These hard thermal loop screening corrections are sufficient to render
finite the damping rate of excitations at rest in the plasma. However these screening corrections are not sufficient
to cure the infrared divergences in the damping rate of charged
excitations at non-zero and large (hard) spatial
momentum\cite{robinfra}. 
The infrared divergences arise from the emission and absorption of
long-wavelength magnetic (transverse) photons or 
gluons which are not screened by the hard thermal loop
corrections\cite{robinfra}.  Whereas longitudinal photons
(instantaneous Coulomb interaction) are screened at finite temperature
with a Debye screening mass $\approx eT$, magnetic 
photons (transverse) are dynamically screened for non zero frequency
as a result of Landau
damping\cite{rob1}-\cite{rob3,blaizot1,robinfra}. However, 
 quasistatic long-wavelength magnetic photons are not screened in the
Abelian theory, and their emission and 
absorption by a fast moving charged particle results in infrared
divergences in the imaginary part of the self-energy on shell. 

These infrared divergences for charged particles are not specific to
particular theories but are somewhat universal in the sense that the
same structure of divergences is common to QED, QCD and scalar QED in
lowest order in the HTL resummation\cite{rate1}-\cite{rebhan}.
Further studies 
of the spectral function questioned the validity of the quasiparticle
approximation and the exponential relaxation associated with a damping
rate\cite{kobes,pilon}. Although these studies provided an
understanding 
of the failure of the quasiparticle picture (exponential relaxation)
for hard fermions, the issue 
of the relaxation time scales was only recently clarified by the
implementation of a Bloch-Nordsieck resummation of the infrared
divergent 
diagrams\cite{iancu,taka} which yields anomalous logarithmic
relaxation. This resummation scheme was previously used 
at finite temperature to verify the cancellation of infrared divergences
of soft photons\cite{welinfra}.  

Infrared divergences in the propagators of charged fields are not particular
of finite temperature field theory. It is well known that electrons in
QED do not 
have a pole associated with their mass shell but rather a cut structure. 
This is a consequence of the emission of soft photons which because of
their masslessness make the 
putative electron mass shell pole to become the beginning of a
cut. This results in that even 
after ultraviolet renormalization, the wave function renormalization
is logarithmically infrared divergent 
on shell. The implementation of the Bloch-Nordsieck resummation of
these infrared 
divergences at zero temperature leads to the correct electron
propagator\cite{bogo,wein}. This resummation at zero temperature is
equivalent to a renormalization group resummation of the leading
infrared divergences in the Euclidean fermion propagator and leads to
an anomalous scaling dimension (albeit gauge  
dependent) for the threshold behavior of fermions in QED\cite{bogo}.

In this article we introduce a {\em dynamical renormalization group} 
resummation programme that allows to obtain the {\em real time} dependence
of retarded propagators, and leads unequivocally to the real time dynamics
of relaxation and thermalization without any assumptions on quasiparticle
structure of collective excitations. This resummation scheme is the
dynamical (real time) equivalent of the renormalization group
resummation  
of divergences in Euclidean Green's functions which is so successful
in both critical phenomena and asymptotic freedom and trascends
approximations 
of the Bloch-Nordsieck type.  The main concept in this programme is the
resummation of secular terms in the perturbative solution of the
equation of evolution of {\em expectation values} that determine the
real time retarded propagators. This dynamical renormalization group
was originally developed to improve the solutions 
of ordinary (and partial) differential equations\cite{goldenfeld}, and
 has been recently implemented in quantum field theory
out of equilibrium\cite{salgado,largen} where it reveals relaxation
with anomalous (and non-perturbative) exponents\cite{largen} (for
other applications in quantum mechanics of few degrees of freedom
see\cite{qm}).  

In the real time description of the dynamical evolution, the time
variable acts as an infrared cutoff. The 
infrared singularities associated with the absorption and emission of
massless quanta are manifest as logarithmic secular 
terms in the perturbative solution of the initial value problem. The
dynamical renormalization group implements  
a non-perturbative {\em resummation} of these secular terms
and leads to anomalous relaxation. In particular for scalar QED we
find that the  charged scalar field expectation value
with hard momentum relaxes in absolute value
as $e^{-\alpha \, T \, t \ln[t/t_0]}$ at asymptotically long times.  The 
asymptotic relaxation is determined by the 
 behaviour of the density of states $ \rho(k;\omega) $
as a function of $ \omega $ near threshold ($ \omega = \sqrt{k^2 + m^2}
$). The larger is $ \rho(k;\omega) $ there, the faster is the decay of the
expectation value of the field. 

The advantage of this method is that it leads to an understanding of
relaxation {\em directly in real time} displaying clearly the contributions
from different regions of the spectral density to the long time behavior. 
Furthermore, it offers a simple criterion to distinguish exponential relaxation and more 
complicated relaxational phenomena
that cannot be interpreted within the quasiparticle picture.

This method implements renormalization group resummations without the
need for invoking a quasiparticle picture or any other approximation.

The article is organized as follows: in section II we provide a direct
link between linear response and the study of relaxation phenomena as
an initial value problem out of equilibrium. In section III we
introduce and test the 
method of dynamical renormalization group within the simple setting
of a field theory of two interacting scalars, one heavy and the other
massless. This simpler theory presents the same type of infrared threshold
singularities as scalar QED and QED. This model presents the infrared
threshold divergences of a critical theory at the upper critical
dimensionality. In this section we compare the
Bloch-Nordsieck approximation and the renormalization group
resummation of infrared divergences in the 
{\em Euclidean} propagator to the real time resummation implemented by 
 the  dynamical renormalization approach in different situations at
zero temperature. This study shows in detail 
the equivalence of all the different approaches at $T=0$. We then implement the
dynamical renormalization group at finite temperature and find anomalous
logarithmic relaxation as in finite temperature QED\cite{iancu}. 
Section IV is devoted to a discussion of the dynamical renormalization
group to elucidate this resummation program and 
 to make contact with the usual renormalization in Euclidena space-time. 
In section
V we study in detail scalar quantum electrodynamics. This theory
has been previously studied within the imaginary time, equilibrium
formulation\cite{rebhan} and shown to have the same type of behavior
as QED and QCD in leading order in the HTL resummation. We study both
the exchange of bare photons and include the HTL  resummation programme consistently to leading order into the
dynamical renormalization group.  This combined resummation of HTL and infrared secular terms in real time
  leads at once to anomalous logarithmic relaxation as in QED in the Bloch-Nordsieck approximation\cite{iancu} and in the simpler scalar case studied in section III.  

We summarize our studies in the conclusion wherein we advocate to use this new approach
based on the dynamical renormalization group to study fermionic excitations
in a plasma and raise further questions and comments. The method of dynamical renormalization group leads directly to an understanding of damping and
relaxation {\em in real time} without invoking a quasiparticle picture or any other approximation. 

Two appendices provide technical details and a third appendix provides a very simple and pedagogical example of the
dynamical renormalization group. 

\section{Preliminaries: from linear response to initial value problem}
We are  interested in studying the real time evolution of expectation
values of  field operators. Consider a scalar field theory with an
interacting Lagrangian density ${\cal L}[\Phi]
$ the expectation value
of the scalar field $\Phi$ can be obtained from linear response to an
external c-number source term $J$. 
The appropriate formulation of real time, non-equilibrium dynamics is that of Schwinger-Keldysh\cite{ctp}-\cite{ata} in which a path 
integral along a contour in
 imaginary time is required to generate all of the non-equilibrium Green's functions.

The non-equilibrium Lagrangian density along this contour 
is therefore given by\cite{ctp}-\cite{ata}
$$
{\cal L}_{NEQ}[\Phi^+,\Phi^-;J] = {\cal L}[\Phi^+]+J\Phi^+ -
{\cal L}[\Phi^-]-J\Phi^-
$$

The non-equilibrium expectation value of the scalar field in a linear
response analysis is given by
$$
\langle \Phi^+(\vec x,t) \rangle   =  \langle \Phi^-(\vec x,t) \rangle = \phi(\vec x,t) = 
\int_{-\infty}^{\infty} d^3{\vec x}' dt' G_R(\vec x- {\vec x}',t-t')
J({\vec x}',t') 
$$
with the retarded Green's function
\begin{eqnarray}
G_R(\vec x- {\vec x}',t-t') &  =  & \left[G^>(\vec x- {\vec x}',t-t')-G^<(\vec x- {\vec x}',t-t')\right]\Theta(t-t') \nonumber \\
& = & i \langle \left[\Phi(\vec x,t), \Phi(
{\vec x}',t' \right]\rangle \Theta(t-t')  \nonumber
\end{eqnarray}
where the expectation value is in the full interacting theory but with
vanishing source. Consider an external source term that is adiabatically
switched on in time from $t \rightarrow -\infty$ and of the form
\begin{equation}
J({\vec x}',t') = J({\vec x}') e^{\epsilon t'} \Theta(-t') \; \; ; \; \; 
\epsilon \rightarrow 0^+ \label{source} 
\end{equation}
The retarded nature of $G_R(\vec x- {\vec x}',t-t')$ results in that
\begin{eqnarray}
\phi(\vec x,t=0) & = & \phi_0(\vec x) \label{inival} \\
\dot{\phi}(\vec x,t<0) & = & 0 \label{inideri}
\end{eqnarray}
where $ \phi_0(\vec x) $ is determined by $J(\vec x)$ (or viceversa,
the initial value $ \phi_0(\vec x) $ can be used to find $J(\vec x)$)
and the vanishing of the derivative for $t<0$ is a consequence of the
retarded nature of $G_R$. 
The linear response problem with the 
initial conditions at $t=0$ given by (\ref{inival})-(\ref{inideri}) can now
be turned into an initial value problem for the {\em equation of motion}
 of the expectation value by using the (integro-) differential operator
${\cal O}_{(\vec x,t)}$ inverse of $G_R(\vec x- {\vec x}',t-t')$
$$
{\cal O}_{(\vec x,t)}\phi(\vec x,t)  =   J(\vec x,t) 
\; \; ; \; \; \phi(\vec x,t=0)  =  \phi_0(\vec x) \; ;
\;\dot{\phi}(\vec x,t<0)  =  0  
$$
for the source term given by eq. (\ref{source}). Within the non-equilibrium
formulation the equation of motion of the expectation value is obtained
via the tadpole method and automatically leads to a retarded initial value
problem by coupling an external source that satisfies eq.(\ref{source}). 

\section{A simple example: a scalar theory}
We begin by considering a simple scalar theory of a massive and a massless
scalar field with Lagrangian density
$$
{\cal L} = \frac{1}{2}(\partial_{\mu}\sigma)^2 + \frac{1}{2}(\partial_{\mu}\pi)^2-\frac{1}{2}m_o^2 \sigma^2 -g_o \sigma^2\pi +J_o\sigma
$$
\noindent where the source coupled to the sigma field has been introduced
to provide an initial value problem as explained in the previous section.
Introducing the following renormalizations
\begin{eqnarray}
&&\sigma= Z^{1/2}_{\sigma}\sigma_r ~\; ; \; \pi=Z^{1/2}_{\pi}\pi_r ~ \; ; \; 
J_r=Z^{1/2}_{\sigma}J_o \nonumber \\
&& m_o^2 Z_{\sigma} = m^2_r Z_{\sigma}+ \delta m^2 ~\; ; \;
g_oZ_{\sigma}Z^{1/2}_{\pi}= g_r Z_g \nonumber 
\end{eqnarray}
\noindent We now suppress the label $r$ with  all quantities being
renormalized, and write the Lagrangian density in terms of
renormalized quantities and counterterms 
\begin{eqnarray}
 {\cal L} & = & \frac{1}{2}(\partial_{\mu}\sigma)^2 +
 \frac{1}{2}(\partial_{\mu}\pi)^2-\frac{1}{2}m^2 \sigma^2 -g
 \sigma^2\pi +J\sigma + {\cal L}_{ct} \nonumber \\ 
 {\cal L}_{ct}&  = &  \frac{1}{2}(Z_{\sigma}-1)
\left((\partial_{\mu}\sigma)^2-m^2\sigma\right) -\frac{1}{2}\delta m^2
 \sigma^2+   
\frac{1}{2}(Z_{\pi}-1)
(\partial_{\mu}\pi)^2-\frac{1}{2}m^2 \sigma^2 -g (Z_g-1)\sigma^2\pi
 \nonumber \\ 
&& -\frac{1}{2}\delta m^2_{\pi} \pi^2\nonumber
\end{eqnarray}
\noindent where we introduced a mass counterterm for the $\pi$ field to 
keep it massless. The counterterms are adjusted in perturbation theory
as usual.

The purpose of studying this simpler model is twofold: i) it provides a
simpler setting to implement and test the method of the dynamical
renormalization group and compare with previous studies of
relaxation\cite{tadpole1,ata}. 
ii) the nature of the infrared divergences in this simpler theory is 
similar to that of gauge theories in lowest order, i.e. the exchange
of a massless field 
in the self-energy of a massive field. These infrared divergences are very
similar to those found in scalar QED\cite{rebhan}, QED\cite{iancu} and
lowest order QCD\cite{lebellac}. After 
studying the resummation of the infrared divergences via the dynamical
renormalization group, we  apply the method to gauge theories. 

Writing $\sigma(\vec x,t) = \psi(\vec x,t)+\phi(\vec x,t)$ with
$\langle \psi(\vec x,t) \rangle =0; \; \phi(\vec x,t)= \langle
\sigma(\vec x,t) \rangle$ using the tadpole condition and taking
spatial Fourier transforms we find the 
equation of motion in the amplitude expansion
$$
\ddot{\phi}_k(t)+(Z_{\sigma}-1)\left[\ddot{\phi}_k(t)+\omega^2_k\;
\phi_k(t)\right]+ 
[\omega^2_k+ \delta m^2]\; \phi_k(t) + \int_{-\infty}^t \Sigma_k(t-t')\;
\phi_k(t')\; dt' =
J_k(t) 
$$
with $\omega^2_k=k^2+m^2$. We have absorbed the contribution of a momentum
and time independent tadpole (ultraviolet and infrared divergent) in a
renormalization of the mass. As discussed in the previous section, the
 source is chosen so that $\phi_k(t=0)=\phi_k(0)\; ; \;
\dot{\phi}_k(t\leq 0) =0$. $\Sigma(t-t')$ is the retarded
self-energy. Writing $\Sigma_k(t-t') = \frac{d}{dt'}\gamma_k(t-t')$
(with the boundary  
condition $\gamma_k(-\infty)=0$ corresponding to adiabatic switching-on of
the interaction) and imposing that $J_k(t>0)=0$, after an integration
by parts the equation of motion for $ t>0 $ becomes
 
\begin{equation}
\ddot{\phi}_k(t)+(Z_{\sigma}-1)\left[\ddot{\phi}_k(t)+
\omega^2_k \;\phi_k(t)\right]+  
\left[\omega^2_k+\delta m^2 + \gamma_k(0)\right] \phi_k(t) - \int_{0}^t
\gamma_k(t-t')\dot{\phi}_k(t')\; dt'=0  \label{fineq} 
\end{equation}

To one loop order we find 
\begin{eqnarray}
\Sigma_k(t-t') = -\frac{g^2}{4\pi^3} \int \frac{d^3 q}{q \omega_{k+q}} &&
 \left\{
(1+N_q+n_{k+q})\sin\left[(q+\omega_{k+q})(t-t')\right]- \right. \nonumber \\
&& \left. (N_q-n_{k+q})\sin\left[(q-\omega_{k+q})(t-t')\right]
 \right\} \label{1lupselfen} 
\end{eqnarray}

\noindent where $N_q$ is the Bose-Einstein distribution function for
the massless field $\pi$ and $n_{k+q}$ is the corresponding
distribution for the massive field $\sigma$. It proves convenient to
write the self energy in the form of a dispersion relation
\begin{eqnarray}
&&\Sigma_k(t-t')  =   \int^{\infty}_{-\infty} d\omega \;
\rho(k;\omega) \;  \sin\left[\omega(t-t')\right] \label{sigmadisp} \\ 
&&\rho(k;\omega)  =  
-\frac{g^2}{4\pi^3} \int \frac{d^3 q}{q \;  \omega_{k+q}} 
 \left\{
(1+N_q+n_{k+q})\delta\left(\omega-q-\omega_{k+q}\right)- 
 (N_q-n_{k+q})\delta\left(\omega-q+\omega_{k+q}\right)\right\} \label{1luprho}
\end{eqnarray}
consequently 
\begin{equation}
\gamma_k(t) = \int \frac{d\omega}{\omega} \rho(k;\omega) \cos(\omega t) \label{gamma}
\end{equation}
 A simple calculation  yields
\begin{eqnarray}
\rho(k;\omega) = -\frac{g^2}{2\pi^2} && \left\{  
\left[ \frac{\omega^2-\omega^2_k}{\omega^2-k^2} + \frac{T}{k}
\ln\frac{ 
(1-e^{-\frac{q^+}{T}})(1-e^{-\frac{\omega-q^-}{T}})}
{(1-e^{-\frac{q^-}{T}})(1-e^{-\frac{\omega-q^+}{T}})}\right\} 
\Theta(\omega -\omega_k) \right. \nonumber \\
&+& \left. \frac{T}{k} \ln
\frac{1-e^{-\frac{q^+}{T}}}{1-e^{-\frac{q^+-\omega}{T}}} \right\}  
\left[\Theta(k-\omega)+ \Theta(k+\omega)\right] \quad ,
 \label{finiteTrho} \\
&& q^{\pm} = \frac{\omega^2 - \omega^2_k}{2(\omega\mp k)}\nonumber
\end{eqnarray}

\noindent There are some noteworthy features of the spectral density (\ref{finiteTrho}) above: i) whereas the zero temperature
contribution vanishes (linearly) near threshold at $\omega =\omega_k$, the finite temperature contribution {\em does not}
vanish at threshold, ii) the finite temperature contribution below the
light cone $ -k < \omega < k$ has its origin 
in Landau damping type processes in which the $\sigma$ particle
scatters off a $\pi$ particle in the medium.  


As it will be seen and understood  below the spectral density
(\ref{finiteTrho}) leads to threshold infrared 
divergences. These will be studied in detail for the cases $T=0
;~T\neq 0$ separately in the next subsections.  


The equation of motion (\ref{fineq}) can be solved by Laplace
transform. In terms of the Laplace transforms of  
$\phi_k(t)$ and $\gamma_k(t)$ given by $\tilde{\phi}_k(s) \; ; \;
\tilde{\gamma}_k(s)$ respectively, with $s$ the Laplace 
transform variable, we find
\begin{equation}
\tilde{\phi}_k(s) = \frac{\phi_k(0)}{s}\left\{1-\frac{\omega^2_k \;  {\cal
C} }{s^2+\omega^2_k+\Pi(s) }\right\} \label{laplaeqn}
\end{equation}
with 
\begin{eqnarray}
{\cal C} & = &  1+(Z_{\sigma}-1)+\left[\delta
m^2+\gamma_k(0)\right]/\omega^2_k \label{calC} \\ 
\Pi(s) & = &  (s^2+\omega^2_k)(Z_{\sigma}-1)+\delta m^2 +
\gamma_k(0)-s \; \tilde{\gamma}_k(s) \nonumber
\end{eqnarray}

The term 
$$
G(k,s) = \left[s^2+\omega^2_k+\Pi(s)\right]^{-1} 
$$
is recognized as the  propagator in terms of the Laplace variable s. The retarded Green's function is obtained by
the analytic continuation
$$
G_{ret}(k,\omega) = G(k,s=i\omega + \epsilon)|_{\epsilon = 0^+} 
$$

The Laplace transform of the self-energy is recognized to be
$$
\tilde{\Sigma}_k(s) = \gamma_k(0)-s \;\tilde{\gamma}_k(s) = \int
d\omega \rho(\omega) \frac{\omega}{s^2+\omega^2}
$$
and its analytic continuation is then given by 
\begin{eqnarray}
\tilde{\Sigma}(s=i\omega+0^+) & = &  \Sigma_R(\omega) + i
\Sigma_I(\omega) \nonumber \\ 
\Sigma_R(\omega) & = &  \int d\omega' \rho(k,\omega') \;  {\cal
P}\frac{\omega'}{\omega'^2-\omega^2} \label{sigmareal} \\
\Sigma_I(\omega) & = & -\frac{\pi}{2}
\left[\rho(k,\omega)-\rho(k,-\omega)\right]\mbox{sign}(\omega)
\label{sigmaimag} 
\end{eqnarray}

\noindent therefore $\Pi(s)$ is recognized as the twice subtracted self-energy which is rendered finite by a proper choice of counterterms. Furthermore, choosing to renormalize at $s^2= - \omega^2_k$ with the
counterterms given by
\begin{eqnarray}
Z_{\sigma}-1 & = &  \frac{\partial \Sigma_R(k,\omega)}{\partial
\omega^2}|_{\omega=\omega_k} \nonumber 
\\ 
\delta m^2 & = & - \int \rho(k,\omega') \; {\cal
P}\frac{\omega'}{\omega^{'2}-\omega^2_k} \; d\omega'\nonumber
\end{eqnarray}
we find 
\begin{equation}
{\cal C} = 1+ \int \frac{\rho(k,\omega')}{\omega'} \; {\cal
P}\frac{\omega^2_k}{(\omega^{'2}-\omega^2_k)^2} \; d\omega'
\label{calCfinita} 
\end{equation}
is {\em finite} even for renormalizable theories in which
$\rho(k,\omega) \sim \omega^2$ 
at large $\omega$ leading to quadratic and logarithmic divergences with
logarithmically divergent wave function renormalizations. This {\em
finite} wave function renormalization will be  seen to emerge naturally from
the dynamical renormalization group.

The real time evolution is obtained by performing the inverse Laplace transform along a path in the complex-s plane
parallel to the imaginary axis to the right of all the singularities of $\tilde{\phi}_k(s)$. We note that the putative
pole at $s=0$ has vanishing residue.

\subsection{Bloch-Nordsieck and Renormalization Group  at $T=0$}

At $T=0$ we find after renormalizing the mass
\begin{eqnarray}
\tilde{\Sigma}(s)=\gamma_k(0)-s\tilde{\gamma}_k(s)& = &
\frac{g^2}{4\pi^2}
\frac{P^2_E+m^2}{P^2_E}\ln\frac{P^2_E+m^2}{m^2}
\label{selflapla}\\ 
P^2_E & = & s^2+k^2 \nonumber
\end{eqnarray}
where $P^2_E$ is identified with the {\em Euclidean} four momentum
 squared. Since in this theory the wave function renormalization is
 finite, we choose $Z_{\sigma}=1$ in what follows. Therefore up to
 this one loop order we obtain the Euclidean irreducible two point
 function (the inverse of  the Green's function) to be given by 
\begin{equation}
\Gamma(k,s)=\left[G(k,s)\right]^{-1} =
m^2_R\left(\frac{P^2_E}{m^2_R}+1\right)\left[1+\frac{g^2}{4\pi^2
P^2_E}\ln\left(\frac{P^2_E}{m^2_R}+1 \right)\right]\label{1lupgam} 
\end{equation}
We clearly see that  $P^2_E \rightarrow -m^2_R$ is no longer a pole
but the end point of a logarithmic branch cut, corresponding to the
threshold for the intermediate state of a $\sigma$ particle and a soft
massless $\pi$ particle.  

This logarithmic infrared divergence near threshold is the same
phenomenon as in gauge theories wherein the self energy 
of charged fields has an infrared logarithmic divergence at threshold
associated with the emission of soft quanta.  
The Bloch-Nordsieck resummation exponentiates the logarithmic
divergences near threshold leading to  
\begin{equation}\label{gBN}
\left. G_{BN}(k,s) = \frac{1}{m^2_R}
\; \left(\frac{P^2_E}{m^2_R}+1\right)^{\lambda-1}
\right|_{\frac{P^2_E}{m^2_R} \approx -1}  
\end{equation}
with the dimensionless coupling
\begin{equation}
\lambda =\frac{g^2}{4\pi^2 m^2_R} \label{lambda}
\end{equation} 
Using this resummed expression for the propagator the inverse Laplace
transform can be performed by wrapping the contour around the branch cuts
along the imaginary axis from $s=\pm i\omega_k$ to $\pm i\infty$. Computing
the discontinuity of $G_{BN}(k,s)$ across these cuts  the real time
evolution of the expectation value is given by
\begin{eqnarray}
\phi_k(t) &=& { \phi_k(0) \over m^2} 
\; \omega^2_k\; {\cal C} \;
m^{2(1-\lambda)}\; 
\frac{2}{\pi}\sin\left[\pi \lambda\right] \;
\int_{\omega_k}^{\infty}\frac{d\omega}{\omega} \frac{\cos(\omega
t)}{\left[\omega^2 - \omega^2_k\right]^{1-\lambda}}  \nonumber
\end{eqnarray}
with $ \omega^2_k = k^2+m^2 $. 

Using now the result\cite{grad}
\begin{eqnarray}
\int_1^{\infty}{ dx \; \cos (x y) \over x \, (x^2 -1 )^{1- \lambda}} &=&
{ \pi^2 \; y \over 4 \, \sin \pi \lambda }\left\{ J_{\frac12 -
\lambda}(y) \; \left[ N_{-\frac12 -\lambda}(y)- {\cal H}_{-\frac12
-\lambda}(y) \right] \right. \cr \cr
&-& \left. J_{-\frac12 - \lambda}(y) \; \left[ N_{\frac12
-\lambda}(y)- {\cal H}_{\frac12 -\lambda}(y) \right]\right\}\nonumber
\end{eqnarray} 
with $ J_{\nu}(z) $ a Bessel function and $ {\cal H}_{\nu}(z) $ a
Struve function, we find the asymptotic long time behavior 
\begin{equation}
\phi_k(t) \buildrel{t \to \infty}\over=
-{ \phi_k(0) \over m^2} 
\; { \omega^2_k {\cal C}\over \Gamma\left(1-\lambda\right)}
\; \left[\frac{m^2}{\omega^2_k}\right]^{1-\lambda}\; 
\left[{2 \over \omega_k\;t}\right]^{\lambda}
\cos\left(\omega_k \; t+\frac{\pi \lambda}{2}\right)
\left[ 1 + {\cal O}\left({1 \over \omega_k \; t}\right)\right] 
\label{asymptotic}
\end{equation}
Here we used the asymptotic formula \cite{grad},
$$
{\cal H}_{\nu}(z) - N_{\nu}(z)\buildrel{z \to \infty}\over= 
{1 \over \sqrt{\pi} \, \Gamma\left(\nu + \frac12\right)} \left( { z
\over 2} \right)^{\nu-1} \left[ 1 + {\cal O}\left({1 \over
z^2}\right)\right] \; .
$$
We see that the Bloch-Nordsieck resummation of the infrared divergences
leads to relaxation with an {\em anomalous exponent}. This is similar to
the case of QED. We now argue that the infrared divergence and the emergence
of anomalous dimensions can be understood by establishing a parallel with
{\em static} critical phenomena at the upper critical dimensionality in
Euclidean space-time. This connection will pave the way to using the renormalization group to sum up infrared divergences non-perturbatively
much in the same manner as in the theory of critical phenomena. In order
to establish this connection more clearly we now introduce the dimensionless
variable
$$
\overline{P}^2 = \frac{P^2_E}{m^2_R}+1 
$$
the main reason for introducing this variable is that when $P^2_E
\rightarrow -m^2_R \; ; \; \overline{P}^2 \rightarrow 0$, therefore
the threshold 
behavior is mapped onto the zero Euclidean four momentum region in terms
of the new variable. In critical phenomena logarithmic divergences appear
when the Euclidean four momentum goes to zero at criticality at the upper
critical dimension. We now introduce a wave function renormalization
constant
$$
Z_{\phi}(\overline{K})= 1+\lambda\ln\overline{K}^2 
$$
and a renormalized irreducible two-point function
$$
\Gamma_R(\overline{P},\overline{K}) = Z_{\phi}(\overline{K})\;
\Gamma(\overline{P}) =  
m^2_R \overline{P}^2 \left[1+\lambda \left(
\frac{1}{\overline{P}^2-1}\ln \overline{P}^2+
 \ln\overline{K}^2\right)\right]
$$
where $\lambda$ is the dimensionless coupling (\ref{lambda}) and
$\overline{K}$ is an arbitrary renormalization point. The bare 
irreducible function $\Gamma(\overline{P})$ is independent of the
renormalization point $\overline{K}$ , i.e. $\overline{K} d\Gamma
/d\overline{K}=0$ which leads to the renormalization group equation 
\begin{eqnarray}
&&\left[\overline{K} \frac{\partial}{\partial\overline{K}}-\eta
\right]\Gamma_R(\overline{K},\overline{P})=0 \label{RGeqn} \\ 
&& \eta= \frac{\partial Z_{\phi}(\overline{K})}{\partial
\ln\overline{K}}=2\lambda  \nonumber
\end{eqnarray}
Near threshold when $\overline{P} \rightarrow 0$ 
$$
\Gamma_R(\overline{K},\overline{P}) = m^2_R~~ \overline{P}^2 ~~
\Phi\left(\frac{\overline{P}}{\overline{K}}\right)
$$
with $ \Phi $ a dimensionless function of its argument. The renormalization
group equation (\ref{RGeqn}) then leads to 
$$
\left[\overline{P} \frac{\partial}{\partial\overline{P}}+\eta
\right]\Phi\left(\frac{\overline{P}}{\overline{K}}\right)=0 
$$
with solution
$$
\Phi\left(\frac{\overline{P}}{\overline{K}}\right)= \Phi(1)
\left[\frac{\overline{P}}{\overline{K}}\right]^{-\eta} 
$$
\noindent finally leading to the renormalization group improved
two-point function 
\begin{equation}\label{resu2p}
G_{RG}(k,s) = \left[ \Phi(1) m^2_R {\overline{K}}^2 \right]^{-1}
\left[\frac{{\overline{K}}^2}{\frac{P^2_E}{m^2_R}+1}\right]^{1-\lambda} 
\end{equation}
which coincides with the one obtained by
the Bloch-Nordsieck resummation, eq. (\ref{gBN}) up to an overall multiplicative factor. 
We can now retrace the same steps that led to the real-time evolution of
the expectation value, by performing the analytic continuation for the
retarded correlation function and the inverse Laplace transform, leading
to the relaxation with anomalous dimension given by eq.(\ref{asymptotic}).

The equivalence between the renormalization
group improved and the Bloch-Nordsieck re-summed propagator is fairly
well known\cite{bogo,wein}. The main purpose of our analysis is to make the point that
the anomalous dimension in the amplitude of the expectation value can be
understood as arising from the scaling behavior of the Green's function 
near threshold in terms of the variable $\overline{P}$. This scaling
behavior, a result of the infrared divergences associated with the
emission of soft quanta is akin to those in static critical phenomena.  

Furthermore, this structure of anomalous dimension of the Euclidean
propagator as a result of threshold infrared 
divergences is similar to that found in QED at $T=0$  either via a
Bloch-Nordsieck or renormalization group  
resummation\cite{bogo,wein}.


In preparation to the forthcoming discussion presented below on the
real time interpretation of the renormalization 
group resummation of infrared divergences, it proves illuminating to
obtain the perturbative form of the real time 
solution. This will allow us to identify the real time manifestation
of infrared divergences. The naive perturbative 
expansion of the renormalization group improved propagator (\ref{gBN})
to first order in the coupling $\lambda$ near 
threshold leads obviously to the one-loop result
(\ref{1lupgam}). After performing the Fourier transform and obtaining 
the real time evolution given by (\ref{asymptotic}) we can now expand
naively in the coupling constant, and find 
\begin{equation}
\phi_k(t) \approx A_k \cos\left(\omega_k t\right)\left[1-\lambda
\ln(\omega_k t)\right] + \cdots \label{firstsec}  
\end{equation}
\noindent with $A_k$ the amplitude read off from  (\ref{asymptotic}). 
This expression reveals that if we attempt a {\em perturbative}
solution of the real-time equation of motion 
(\ref{fineq}) we would find {\em logarithmic secular terms},
i.e. terms that grow in time and invalidate the 
perturbative solution at long times. In this case a perturbative
expansion of the real-time equation of motion will break down 
at time scales $t_{break} \approx  e^{1/\lambda}/\omega_k$. The
renormalization group in the energy representation 
provides a resummation of the infrared divergences in the propagator,
which  leads to a real-time evolution that is 
asymptotically decreasing function of time.

 We now study the perturbative solution of  (\ref{fineq}) that reveals
indeed these secular terms, and implement a real-time version of the
renormalization group that implements precisely this 
resummation. 


\subsection{Dynamical Renormalization Group at $T=0$}

Having established the resummation of threshold infrared divergences both
within the Bloch-Nordsieck approximation and the renormalization group, we
now introduce a novel method that allows a similar resummation but {\em 
directly in real time}. Consider seeking a solution of the equation of
motion (\ref{fineq}) in perturbation theory in the coupling. Writing
the self energy as an expansion in terms of the dimensionless coupling
$\lambda$ given by (\ref{lambda}), $\Sigma_k = \sum_{n=1}\lambda^n
\Sigma_k^{(n)}$ a perturbative solution obtained as a power series 
expansion is given by $\phi_k(t)=\phi^{(0)}_k(t)+\lambda
\phi^{(1)}_k(t)+\cdots$ with the hierarchy of equations 
\begin{eqnarray}
\ddot{\phi}^{(0)}_k(t)+ \omega^2_k \phi^{(0)}_k(t) & = & 0 \nonumber \\
\ddot{\phi}^{(1)}_k(t)+ \omega^2_k \phi^{(1)}_k(t) & = & -\left[\delta
m^2+\gamma^{(1)}_k(0)\right]\phi^{(0)}_k(t) + \int_0^t
\gamma^{(1)}_k(t-t')\; \dot{\phi}^{(0)}_k(t')\; dt' \nonumber \\ 
\quad \vdots \quad \quad \quad & = & \quad \quad \quad  \vdots \quad
\nonumber
\end{eqnarray}
where we note that the contribution from the wave-function
renormalization vanishes by virtue of the zeroth-order equation of
motion. The solution to the zeroth-order equation is  
$$
\phi^{(0)}_k(t) = A_k e^{i\omega_k  t}+A^*_k e^{-i\omega_k  t}
$$
and the initial conditions $\phi_k(t=0)=\phi_k(0); \; ; \dot{\phi}_k(t=0)=0$ implies $A_k=A^*_k=\phi_k(0)/2$, but we
will leave both constants to recognize more easily the different contributions and we will  use this condition at the
end of the calculations. The 
solution to the above hierarchy of  equations can be found in terms of the 
retarded Green's function of the unperturbed problem
$$
{\cal G}_R(t_1-t_2) = \frac{1}{\omega_k}\sin[\omega_k (t_1-t_2)]\Theta(t_1-t_2)
$$
The higher order corrections are easily (but tediously) computed using
the spectral representation of the self-energy. The first order term
is given by
\begin{eqnarray}
\phi^{(1)}_k(t) & = & \phi^{(1,a)}_k(t)+\phi^{(1,b)}_k(t) \nonumber\\
\phi^{(1,a)}_k(t) & = & \frac{1}{\omega_k} \int \frac{d\omega}{\omega}
\rho(\omega) \int_0^t dt' \int_0^{t'}dt_1\;
\sin[\omega_k(t-t')]\cos[\omega(t'-t_1)\;\dot{\phi}^{(0)}(t_1)
\nonumber \\ 
\phi^{(1,b)}_k(t) & = & -\frac{1}{\omega_k}\left(\delta m^2 + \int
\frac{d\omega}{\omega} \rho(\omega)\right) \int_0^t
\sin[\omega_k(t-t')]{\phi}^{(0)}(t')\;dt' \nonumber 
\end{eqnarray}
where the spectral density $\rho(\omega)$ is given to one loop order by
eq.(\ref{1luprho}) and its leading infrared contribution for $T \neq 0$ given by eq.(\ref{finiteTrho}). The integral over the time variables
can be done straightforwardly,  the result is given by: 
\begin{eqnarray}
\phi^{(1,a)}_k(t) = && -\frac{i}{4} \int
\frac{d\omega}{\omega}\;\rho(\omega)\left\{  
A_k e^{i\omega_k t} \left[\frac{1}{\omega_k- \omega}\left(t-
\frac{e^{i(\omega-\omega_k)t}-1}{i(\omega-\omega_k)}\right) + \omega
\rightarrow -\omega \right]- \right. \nonumber \\ 
&& \left. A_k e^{-i\omega_k t}\left[\frac{1}{\omega_k- \omega}\left(
\frac{e^{2i\omega_k
t}-1}{2i\omega_k}-\frac{e^{i(\omega_k+\omega)t}-1}{i(\omega_k+\omega)}\right)
+ \omega \rightarrow -\omega \right]+ \right.\nonumber \\ 
&& \left. A^*_k e^{i\omega_k t}\left[\frac{1}{\omega_k+ \omega}\left(
\frac{e^{-2i\omega_k t}-1}{-2i\omega_k}+\frac{e^{-i(\omega_k-\omega)t}-1}{i(\omega_k-\omega)}\right)+ \omega \rightarrow -\omega \right]- \right. \nonumber \\
&& \left. A^*_k e^{-i\omega_k t} \left[\frac{1}{\omega_k+ \omega}\left(t-
\frac{e^{i(\omega+\omega_k)t}-1}{i(\omega+\omega_k)}\right) + \omega
\rightarrow -\omega \right]\right\} \label{fi1atime}  
\end{eqnarray}
\begin{eqnarray}
\phi^{(1,b)}_k(t) = && \frac{i}{2 \omega_k} \left(\delta m^2 +\int
\frac{d\omega}{\omega}\rho(\omega)\right) \left\{  
A_k \; e^{i\omega_k t} \; t - A^*_k \; e^{-i\omega_k t} \; t
\right. \nonumber \\ 
&-& \left. A_k\; e^{-i\omega_k t} \;\frac{e^{2i\omega_k t}-1}{2i\omega_k} -
 A^*_k \; e^{i\omega_k t} \;\frac{e^{-2i\omega_k t}-1}{2i\omega_k}
\right\} \label{fi1btime} 
\end{eqnarray}
Secular terms will arise from  the  contributions of the form
$ e^{i(\omega-\omega_k)t}-1 $ if the  
coefficients of these terms produce singularities in the integration region. 

We are now in condition to analyze  different cases. Although we are
primarily interested in applying the method 
of dynamical renormalization group to the situation of infrared
divergences, in order to gain insight and test this 
method we begin by studying situations in which results are known. To
this purpose we address the familiar case of a 
 generic interacting scalar theory in which the pole
frequency $\omega_k$ is away from thresholds ($\omega_{th}$), either
above, in which 
case there is a resonance,  or below in which case the particle is stable.  

\subsubsection{$\omega_k \neq \omega_{th}$}
For $\omega_k < \omega_{th}$ there are no singularities in the
integration region, therefore the only secular 
terms are those linear in time in $\phi^{(1,a)}_k(t)\; ; \; \phi^{(1,b)}_k(t)$. If on the other hand
 $\omega_k > \omega_{th}$ and far away from threshold, there are singularities (simple and double poles) in the
integration region. 

We can extract the
secular terms in the above expression in both cases $\omega_k > \omega_{th}$ and $\omega_k < \omega_{th}$ by using the results of appendix B or alternatively  taking the 
long time limit using the distributions (see appendix B)
\begin{eqnarray}
\stackrel{\mbox{lim}}{t \rightarrow \infty}
\frac{1}{\alpha^2}\left[\alpha t -{\sin\alpha t}\right]
& = & {\cal P}\left(\frac{1}{\alpha^2}\right)\left[\alpha
t-{\sin\alpha t}\right] \nonumber \\ 
\stackrel{\mbox{lim}}{t \rightarrow \infty} \frac{1-\cos\alpha
t}{\alpha^2} 
& = & \pi t \delta(\alpha)+{\cal
P}\left(\frac{1}{\alpha^2}\right)\left(1-\cos\alpha t \right) 
\nonumber 
\end{eqnarray}
where the $\delta(\alpha)$ accounts for resonant denominators.This term is
recognized from the familiar Fermi's Golden rule.  

Gathering the secular terms from both contributions
(\ref{fi1atime},\ref{fi1btime}) and taking $A_k = A^*_k$ we find 
\begin{eqnarray}
\phi^{(1)}_k(t) = A_k e^{i\omega_k t} &&\left\{
\left[i \frac{\delta m^2+ \Sigma_R(\omega_k)}{2\omega_k} t -
\frac{\Sigma_I(\omega_k)}{2\omega_k} t \right]+ \int
\frac{d\omega}{\omega} \; 
\rho(\omega) \; {\cal
P}\frac{\omega^2_k}{(\omega^2_k-\omega^2)^2}
\right. \nonumber \\ 
 && \left. -\int \frac{d\omega}{\omega}\; e^{i\omega t}\;
\rho(\omega) \; {\cal P}\frac{\omega^2_k}{(\omega^2_k-\omega^2)^2} \right\} 
 + \mbox{c.c}  
\label{fi1sec}
\end{eqnarray}
where  we have used the expressions for the real and imaginary parts of
the analytically continued self-energy given by equations
(\ref{sigmareal})-(\ref{sigmaimag}). The last terms are non-secular at
long times and remain perturbatively small.  

In this manner the resummation of the secular terms is obvious and
correspond to a shift in the pole position $ \omega_k \rightarrow
\omega_k+\delta \omega_k $ (finite by a proper choice of $ \delta m^2 $)  
and a width or decay rate $ \Gamma_k $ given by the imaginary and real part
respectively of the first order correction above. Since this is the 
simplest and most familiar setting to introduce the dynamical
renormalization group, we now present the resummation of the secular terms 
via this method. This is achieved by  introducing a (complex)
renormalization of the amplitude and writing\cite{goldenfeld,salgado} 
\begin{eqnarray}
A_k & = &  {\cal A}_k(\tau) \; {\cal Z}_k(\tau)\label{amplitude} \\
{\cal Z}_k(\tau) & = & 1+\lambda z^{(1)}_R(\tau)+ i \lambda z^{(1)}_I(\tau) +\cdots
\label{zetadyn}
\end{eqnarray} 
where $\tau$ is an arbitrary time scale that acts as a renormalization
point and the $z^{(n)}_{R,I}(\tau)$ are real functions. Choosing 
$$
\lambda z^{(1)}_R(\tau) = \frac{\Sigma_I(\omega_k)}{2\omega_k} \; \tau
\; ; \; \lambda z^{(1)}_I(\tau) =  
-\frac{\delta m^2 + \Sigma_R(\omega_k)}{2\omega_k} \; \tau 
$$
we obtain 

\begin{equation}
\phi_k(t) = {\cal A}_k(\tau) e^{i\omega_k t}\left[1+ i \frac{\delta
m^2+\Sigma_R(\omega_k)}{2\omega_k}\; (t-\tau) -
\frac{\Sigma_I(\omega_k)}{2\omega_k}\; (t-\tau) \right] + \mbox{c.c} +
\mbox{regular terms} 
\label{firstorderren}
\end{equation} 
where the regular terms refer to the non-secular last term  
The meaning of the above expression is clear: a change in the time
scale corresponds to a change in the (complex) amplitude of the expectation
value. Whereas the original perturbative expansion was only valid for
times such that the contribution from the secular terms remain very small compared to the unperturbed
value, the renormalized expression (\ref{firstorderren}) remains
valid for intervals $t-\tau$ such that the secular terms remain small. By
choosing $\tau$ arbitrarily close to $t$ we have improved the
perturbative expansion. 
However $\phi_k(t)$ does not depend on $\tau$: a change of the renormalization point $\tau$ is compensated by a change in the complex
amplitude ${\cal A}_k$. This  leads to the {\em dynamical 
renormalization group equation} to lowest order
$$
\frac{\partial {\cal A}_k(\tau)}{\partial \tau} -\left[i \frac{\delta
m^2+\Sigma_R(\omega_k)}{2\omega_k}  -
\frac{\Sigma_I(\omega_k)}{2\omega_k}  \right]{\cal A}_k(\tau) =0
$$
with obvious solution
\begin{eqnarray}
{\cal A}_k(\tau) & = &  {\cal A}_k(0) \;e^{i \delta \omega_k \tau} \;
e^{-\Gamma_k \tau} \nonumber \\
\delta \omega_k & = & \frac{\delta m^2+\Sigma_R(\omega_k)}{2\omega_k}
\nonumber \\ 
\Gamma_k & = & \frac{\Sigma_I(\omega_k)}{2\omega_k} \nonumber
\end{eqnarray}

Choosing the arbitrary scale $\tau$ to coincide with the time $t$ we obtain
the {\em resummed} expression for the expectation value
\begin{eqnarray}
\phi_k(t) & = &  {\cal C} {\cal A}_k(0)\; e^{i \omega_p(k) t}\;
e^{-\Gamma_k t}-{\cal A}_k(0) \int \frac{d\omega}{\omega} \; e^{i\omega t}\; 
\rho(\omega) \;{\cal P}\frac{\omega^2_k}{(\omega^2_k-\omega^2)^2}
+ \mbox{c.c} \label{resu} \\ 
{\cal C} & = & 1+ \int \frac{d\omega}{\omega}\;
\rho(\omega) \;{\cal
P}\;\frac{\omega^2_k}{(\omega^2_k-\omega^2)^2}\nonumber
\end{eqnarray}
with $\omega_p(k)$ the pole position shifted by one-loop corrections and
$ \Gamma_k $ is identified with the decay or damping rate. The constant
 $ {\cal C} $ is the same as in eq. (\ref{calCfinita}). The on-shell
renormalization leading to eq. (\ref{calCfinita}) 
is here a consequence of the perturbative expansion in terms of the
solutions of the equations of motion.  

After some straightforward algebra, the constant ${\cal C}$ is  found
to be the same as the residue of the Laplace  
transform \ref{laplaeqn}) at the pole (or resonance) at
$\omega_p$. The last, non-secular terms in \ref{resu},  
allows us to 
make contact with previous results\cite{ata}. The long time dynamics
of this integral is dominated by the 
threshold contribution\cite{ata}. If the spectral density vanishes
near threshold as $\rho(\omega) \approx (\omega-\omega_{th})^{\alpha}$
then the asymptotic time evolution is described by a power law
relaxation $t^{-\alpha-1}$ (long 
time tails). Thus we see that the dynamical renormalization group
resummation has obtained all of the features of  
the solution via the Laplace transform (\ref{laplaeqn}) which were
previously obtained\cite{ata}.

The resummation
via the dynamical renormalization group has led to a (asymptotic) convergent
perturbative expansion for the time evolution of the expectation value.

This simple case provides for a clear understanding of the resummations
implied by the dynamical renormalization group and paves the way for
understanding the more complicated cases of threshold singularities and
finite temperature below. 

\subsubsection{ Threshold singularities} 

Having established the reliability of the dynamical renormalization
program in more familiar settings, we are now 
in position to apply this method to study the case of threshold
infrared divergences arising from the emission of soft massless
quanta. Thus we now return to the theory of a massive and a massless
scalar fields of the beginning of this 
section. We begin by analyzing the situation at $T=0$ to make contact 
with the Bloch-Nordsieck and Euclidean 
renormalization group resummations, but now implementing the {\em
dynamical renormalization group} resummation. 

 Since near threshold the spectral density (\ref{finiteTrho}) for
 $T=0$ becomes ($\lambda$ is defined by eq. (\ref{lambda})) 
$$
\rho(k,\omega) \buildrel{\omega \to \omega_k}\over=
-4 \, \lambda \; \omega_k \; (\omega-\omega_k) + {\cal
O}(\omega-\omega_k)^2\; , 
$$
besides the linear secular terms in $\phi^{(1,a)}_k ; \phi^{(1,b)}_k$
(see equations (\ref{fi1atime})-(\ref{fi1btime}) there 
is a potentially infrared divergent term arising from  $\phi^{(1,a)}_k
$ which is {\em real} and using the results of appendix B, found to be
given by  
$$
\frac{1}{4}\int \frac{d\omega}{\omega}\rho(k;\omega)
\frac{1-\cos[(\omega-\omega_k)t]}{(\omega - \omega_k)^2}
\buildrel{\mu t >> 1 }\over= -\lambda (\ln{\mu t}+\gamma) +{\cal O}(\lambda)
$$
where $ \gamma $ is  Euler-Mascheroni constant (see details in
appendix B). We note 
that time acts as an infrared cutoff in the sense that for $\omega \approx
\omega_k$ at finite time the integral is convergent. The infrared divergences
are now manifest in a logarithmic time dependence. The linear secular
terms combine just as in the previous case to provide an {\em imaginary}
secular term given by $i {(\delta
m^2+\Sigma_R(\omega_k))}/{2\omega_k}$ just as in eq.(\ref{fi1sec})
which in this case is simply   frequency shift as can be seen from the
expression for the self-energy given by eq.(\ref{selflapla}). This
shift is made finite with a proper choice of $\delta m^2$.  Thus in
this case we find 
$$
\phi_k(t) = A_k e^{i\omega_k t}\left[1+ i \frac{\delta m^2+
 \Sigma_R(\omega_k)}{2\omega_k}\;  t  - \lambda
\ln\overline{\mu} t \right] + \mbox{c.c} + \mbox{regular terms}
$$
with $\overline{\mu}=\mu e^{\gamma}$.
Similarly to the previous case, we introduce the complex amplitude
(\ref{zetadyn}) and choose 
$$
\lambda z^{(1)}_R(\tau) = \lambda \ln\overline{\mu} \tau \; ; \;
\lambda z^{(1)}_I(\tau) =  
-\frac{\delta m^2+\Sigma_R(\omega_k)}{2\omega_k}\; \tau = -\delta
\omega_k\;  \tau 
$$
leading to the following expression for $\phi_k(t)$
$$
\phi_k(t) = {\cal A}_k(\tau)\; e^{i\omega_k t}\left[1+ i \delta \omega_k
(t-\tau) - \lambda \ln\frac{t}{\tau} \right] + \mbox{c.c} +
\mbox{regular terms} 
$$
 $\phi_k(t)$ is independent of the arbitrary time scale $\tau$ leading
to a renormalization group equation obeyed by the complex amplitude, which
is now given this  order by
$$
\frac{\partial {\cal A}_k(\tau)}{\partial \tau} -\left[i \delta \omega_k  - \frac{\lambda}{\tau} \right]{\cal A}_k(\tau) =0 
$$
with  solution
$$
{\cal A}_k(\tau)  =   {\cal A}_k(\tau_0) \;e^{i\; \delta \omega_k
\tau}\left[\frac{\tau_0}{\tau}\right]^{\lambda}  
$$
Again, choosing the scale $\tau$ to coincide with the time $t$, we finally
obtain the asymptotic dynamics of the expectation value in this
case to be given by
\begin{equation}\label{ampliesc}
\phi_k(t) = {\cal A}_k(\tau_0)\; e^{i \omega_{k,R} t}
\left[\frac{\tau_0}{t}\right]^{\lambda} +\mbox{c.c} +\mbox{small} 
\end{equation}
where $\omega_{k,R}=\omega_k+\delta \omega_k$. The terms denoted by
small remain perturbative at all times and decay faster than 
the term with the anomalous dimension in weak coupling. This
expression coincides with the 
long time behavior found in the previous sections via the
Bloch-Nordsieck and the renormalization group resummation of the
logarithmic infrared divergences of the propagator given by
eq.(\ref{asymptotic}). We thus conclude that the  dynamical 
renormalization group implements a resummation in {\em real time} which
is complementary to the renormalization group or Bloch-Nordsieck
resummations in the frequency representation of the propagator.

\subsection{ $T \neq 0$: Dynamical Renormalization Group resummation}

At finite temperature the infrared divergences are enhanced by the
Bose-Einstein distribution function of massless  
particles $N_q \approx T/q \; ; \; q/T <<1$. This can be seen at the
level of the spectral density $\rho(k;\omega)$  
given by eq.(\ref{finiteTrho}). Whereas the zero 
temperature contribution vanishes linearly near threshold, the finite
temperature contribution remains constant there provided $m \neq
0$. In particular we 
find that near threshold the Laplace transform of the retarded propagator
behaves as
\begin{eqnarray}
&&\Gamma(k,s;T)=\left[G(k,s;T)\right]^{-1} =
\left({P^2_E}+{m^2_R}\right)\left[1+ 
\frac{g^2}{4\pi^2 P^2_E}\ln\left(\frac{P^2_E}{m^2_R}+1
\right)\right]-\overline{g}(T,k)\ln\frac{P^2_E+m^2_R}{\mu^2}\nonumber\\
&&\overline{g}(T,k)= \frac{g^2}{4\pi^2}\left(\frac{T}{k}\right)\ln
\frac{\omega_k+k}{\omega_k-k} \nonumber 
\end{eqnarray}

\noindent where $\mu$ is an arbitrary infrared cutoff scale and
assumed that the external momenta is 
not too hard.  Thus whereas the zero temperature 
inverse propagator actually vanishes at threshold (with an infrared
divergent slope) the finite temperature propagator diverges there,
reflecting the stronger infrared divergence at finite temperature. In
this 
situation a Bloch-Nordsieck resummation of the Euclidean propagator is
not clear as was emphasized in reference\cite{iancu} and a
multiplicative 
(wave function) renormalization cannot cure the infrared divergence since
the finite temperature part is {\em not} proportional to  ${P^2_E+m^2_R}$.

It is precisely in this situation that the power of the dynamical
renormalization group is revealed. We will study two different cases 
in detail: i) hard external momentum $k >>m$ (or $m=0$) and ii) soft
external momentum $k \leq m$. 

\subsubsection{Hard external momentum ($m \approx 0$)}

In this case the Landau damping cut coalesces with the cut for $\omega
\geq k$ and both terms of the spectral density (\ref{finiteTrho})
contribute. 
Because of the simmetry $\omega \rightarrow -\omega$ of the time dependent
terms in eq.(\ref{fi1atime}) the frequency integral in the interval 
$-k < \omega <k$ plus the integral in the interval $k<\omega <\infty$
can be folded into an integral in the range  
$0< \omega <\infty$ in terms of the effective 
finite temperature spectral density
$$
\overline{\rho}(k;\omega) = -\frac{g^2}{\pi^2}\frac{T}{k} 
\ln\left[\frac{1-e^{-\frac{|\omega+k|}{2T}}}{1-e^{-\frac{|\omega-k|}{2T}}}\right]\Theta(\omega)
$$
The asymptotic time dependence is dominated by the region $\omega \approx k$
which gives the infrared divergences of the propagator. In this region
the effective spectral density 
\begin{equation}
\overline{\rho}(k;\omega)  \buildrel{\omega \to k}\over=
\frac{g^2}{\pi^2}\frac{T}{k} 
\ln\frac{|\omega-k|}{2\overline{T}_k} \;\Theta(\omega)+ {\cal
O}(\omega-k) \; \; ; \; \; 
\overline{T}_k = \frac{T}{1-e^{-\frac{k}{T}}} \label{effspecscal}
\end{equation}
We start by analyzing the different contributions to  the infrared behavior of the coefficient of $A_k e^{i\omega_k t}$.
A simple analysis leads to the following conclusions: i) the contribution near threshold to the {\em imaginary} part 
of the
coefficient cancels out between the production cut ($\omega >k$) and the Landau-damping cut ($0< \omega < k$) leaving a 
linear secular term without
infrared divergences that renormalizes the mass. ii) The contribution
near threshold to the {\em real} part is the same for both cuts $0 < \omega < k$ and $\omega > k$ and add
up. This contribution in the asymptotic long time limit is obtained from the formulae in appendix B and given by
\begin{equation}
\frac{g^2}{\pi^2}\frac{T}{2k^2}\;  t \int^{\infty}_0 \frac{dz}{z^2}\ln\left[
\frac{z}{2t\overline{T}}\right](1-\cos z ) =  -\frac{g^2}{4\pi k^2} \;
T \; t \; \ln\overline{\mu}t \; \; ; \; \; \overline{\mu} =
2\overline{T}\; e^{\gamma-1} \label{longtime} 
\end{equation}
\noindent where we have quoted the leading contribution in the
asymptotic time regime. Subleading terms can be 
consistently obtained using the formulae of appendix B. 
The coefficient of $ A^*_k e^{i \omega_k t} $ has a potential infrared
divergence, 
however the contribution from both cuts cancel each other leaving an
infrared (and ultraviolet) finite 
result without secular terms. The zero tempreature contribution and that
of $\phi^{(1,b)}_k$ lead to a finite frequency shift. Implementing the
dynamical renormalization group resummation we find
$$
\phi_k(t) = {\cal A}_k(t_0) \;e^{i\omega_{k,R}(t-t_0)} \; e^{-a_k \, T \, t
 \,\ln[t/t_0]}+ \mbox{c.c.}  
~ \; \; ; \; \; a_k= \frac{g^2}{4\pi k^2} 
$$
\noindent where we have solved the renormalization group equation with an
initial condition at a time $t_0 = 1/\overline{\mu}$. This solution reveals
clearly the renormalization group invariance, a change in the arbitrary
time $t_0$ is compensated for by a change in the amplitude and an
overall phase.   

\subsubsection{Soft external momentum ($m \neq 0$)}

For $m \neq 0$ the Landau damping cut and the production cut are
separated and the infrared divergences arise only from 
the production cut $\omega_k < \omega$. In the high temperature limit
we find near threshold  
\begin{eqnarray}\label{softrho}
\rho(k;\omega) &\buildrel{\omega \to \omega_k}\over=&
-\frac{g^2}{\pi^2} \; \frac{T}{k}\;
\ln{\cal M}_k
\left[1+{\cal O}(\omega-\omega_k)\right] \Theta(\omega-\omega_k)\cr\cr
{\cal M}_k & \equiv & 1 + {2\, k \over m^2}(k + \omega_k) \; .
\end{eqnarray}

 We note that in this case the spectral density $\rho(k;\omega)$
 approaches a constant value at threshold.  

The terms proportional to $t/(\omega + \omega_k)$ in
$\phi^{(1,a)}_k(t)$ do not have infrared divergences but they remain 
as secular terms that combine with those of $\phi^{(1,b)}_k(t)$ to
give a renormalization of the frequency just as in the previous
cases. 
The infrared divergences arise from terms with denominators $1/(\omega
- \omega_k)$.  


In this case these infrared divergences are manifest as logarithmic
secular terms in the real and imaginary parts leading 
to damping and anomalous logarithmic phases. 


These contributions are the following: i) the {\em imaginary} part of
the coefficient of $A e^{i\omega_kt}$ is given asymptotically by 
(see appendix B)
\begin{eqnarray}
&&\frac{i\;  t}{4} \int_{\omega_k}^{\infty} d\omega \;
\frac{\rho(k;\omega)}{\omega} \frac{1}{\omega-\omega_k}\left[1- 
\frac{\sin(\omega-\omega_k)t}{(\omega-\omega_k)t}\right]
\buildrel{\mu t >> 1}\over=-  \frac{ i \, g^2\, \ln {\cal M}_k}{4\pi^2\;
k \; \omega_k} \;  T \; t \ln[\mu t\, e^{\gamma-1}] \cr \cr
&&+ i\; t \int_{\omega_k}^{\infty} {d\omega  \over \omega - \omega_k}\left
[ \frac{\rho(k;\omega)}{\omega} -
\frac{\rho(k;\omega_k)}{\omega_k}\Theta(\mu-\omega+\omega_k) \right]
+ {\cal O}\left( {g^2 \over t } \right)
\label{imagina} 
\end{eqnarray}
Notice that eq.(\ref{imagina}) is {\bf independent} of the scale $ \mu $ as
one can easily see since  the derivative with respect to $ \mu $ of
the r. h. s. identically vanishes. The scale $ \mu $ has been
introduced just to have a dimensionless argument in the logarithms.
ii) The {\em real} part is asymptotically given by 
(see appendix B for details) 

\begin{eqnarray}
&& \frac{1}{4} \int_{\omega_k}^{\infty} d\omega \;
\frac{\rho(k;\omega)}{\omega} 
\left[1-\cos(\omega-\omega_k)t\right]\left[\frac{1}{(\omega-\omega_k)^2}-
\frac{1}{2\omega_k(\omega-\omega_k)}\right]
\buildrel{\mu t >> 1}\over= \nonumber \\ 
&&   -\frac{g^2 \, T \; t}{8\,\pi\, k\, \omega_k} \ln {\cal M}_k\; -
\frac{g^2\, T}{4\,\pi^2\, m^2\,
\omega^2_k}\left[1+ {3\, m^2 \over 2\, k\, \omega_k}  \ln {\cal M}_k
\right]\ln[\overline{\mu}\,t\, e^{\gamma}] + {\cal O}(g^2) \nonumber
\end{eqnarray}

The remaining secular but infrared safe terms in $\phi^{(1,a)};\phi^{(1,b)}$
contribute to the imaginary part a term proportional to $t$ which can
be absorbed in a redefinition of the arbitrary scale $\mu$ in
eq. (\ref{imagina}).  The logarithmic divergence as $k/m\rightarrow \infty$
reflects precisely the logarithmic time dependence found 
in the previous case of hard momentum $m \approx 0$ and that results in
the anomalous relaxation proportional to $t \ln \mu t $ as given by
eq. (\ref{longtime}).

Implementing a resummation of the secular terms with the dynamical
renormalization group following the steps outlined above leads to the
asymptotic form 
\begin{eqnarray}\label{escasuave}
\phi_k(t) & = &  {\cal
A}_k(t_0)e^{i\varphi_k(t,t_0)}\left[\frac{t_0}{t}\right]^{\overline{b}_k} 
e^{-\overline{\Gamma}_k t} + \mbox{c.c.} \nonumber \\
\varphi_k(t) & = & \omega_{k,R} \;(t-t_0) + \frac{  g^2\, \ln{\cal
M}_k}{4\pi^2 k \omega_k}  \, T \; t \; \ln{t\over t_0}
\nonumber \\  
\overline{\Gamma}_k & = & \frac{g^2 \; T}{8\pi k \omega_k}
\ln{\cal M}_k \nonumber \\ 
\overline{b}_k & = & \frac{g^2T}{4\pi^2 m^2\, \omega_k}\left[ 1 + {3\,
m^2 \over 2\, k \,\omega_k}  \ln {\cal M}_k \right]\; . 
\end{eqnarray}

It is interesting to try and understand the exponential damping in this
case. For this we show in figure 1 the spectral density for the sigma
field in this soft case near threshold 
\begin{eqnarray}
S(k,\omega;T) & = &
\frac{\Sigma_I(\omega,k;T)}{\left[\omega^2-\omega_k^2
-\Sigma_R(\omega,k;T)\right]^2+\left[\Sigma_I(\omega,k;T)\right]^2}\nonumber
\\ 
\Sigma_R(\omega,k;T) & = & -\frac{1}{2}\rho(k,\omega_k) \; 
\ln\frac{\omega^2-\omega_k^2}{\mu^2}+{\cal O}(\omega-\omega_k)  \nonumber \\ 
\Sigma_I(\omega,k;T) &= & -\frac{\pi}{2} \rho(k,\omega_k)
+{\cal O}(\omega-\omega_k) \label{sigmaspec} 
\end{eqnarray}
 where $\rho(k,\omega_k)$ is given by eq.(\ref{softrho}) and for the figure
it was taken to be $\rho(k,\omega_k)=-0.005$. 
 
 \noindent $S(k,\omega;T)$ vanishes at threshold but with a singular
slope, this results in that  the spectral density features a sharp
peak near threshold, which 
is found to be at $\omega \approx
\omega_k+\frac{1}{2}\rho(k,\omega_k)\ln|\rho(k,\omega_k)|+\cdots$. 

 The decay rate
is given by $ (1/2) \Sigma_I(\omega_k)/(2\omega_k) $ the extra factor
$ 1/2 $ is simply 
a result of the fact that $\omega_k$ is the threshold, i.e. the
end-point of the integral and therefore the on-shell delta
function only picks-up half of the contribution. The logarithmic
dependence of the phase is a consequence 
of the logarithmic infrared threshold divergences and prevents an
interpretation in terms of quasiparticle poles.

This case must be contrasted to that of $m=0$, wherein the imaginary
part of the self-energy is {\em infrared singular} at threshold. This
is revealed in the logarithmic singularity in the limit $ k/m\rightarrow
\infty $ which reflects precisely the logarithmic time dependence found 
in the previous case of hard momentum $m \approx 0$ leading to
the anomalous relaxation proportional to $t \ln[\mu t]$ as given by
eq. (\ref{longtime}).


\section{Discussion and Interpretation:}

Before proceeding further to the case of a gauge theory, it is
convenient to pause and analyze the results that we 
have obtained so far and elucidate the main aspects of the dynamical
renormalization group. The scalar model chosen 
in the previous section is a  non-trivial example of a
superrenormalizable theory that displays the same type of infrared
divergences as a critical theory at the upper critical dimension. The
usual renormalization group leads to a Bloch-Nordsieck resummation and
anomalous dimensions at $T=0$.  
Performing the Fourier transform we recognized that the real-time
interpretation of the renormalization group resummation corresponds to
a power law relaxation with the power determined by the anomalous
dimension. 
A naive perturbative expansion in the dimensionless coupling results in 
secular terms, i.e. terms that grow in time and signal the breakdown of
the perturbative expansion at long time scales. These same secular
divergences are obtained directly in real time when the equation of
motion for the expectation value is solved in a perturbative
expansion. The dynamical renormalization group implements a
resummation of these secular divergences which leads to an improved
perturbative solution.  
The renormalization procedure can be understood with a very simple
and pedagogical example, the weakly damped harmonic oscillator. Consider
the equation of motion
$$
\ddot{y}+y=-\epsilon \dot{y}~, \; \; \epsilon <<1 \label{damposc}
$$
attempting to solve this equation in a perturbative expansion in
$\epsilon$ leads to the lowest order solution (see appendix C) 

$$
y(t)= A\; e^{it}\left[1-\frac{\epsilon}{2}\; t\right] + \mbox{c.c}
$$
\noindent where the term that grows in time, i.e. the linear secular term
leads to the breakdown of the perturbative expansion at time scales
$t_{break} \propto 1/\epsilon$. The dynamical renormalization
introduces a time scale $\tau$ in the form 
$ A=A(\tau) \; Z(\tau) \; ; Z(\tau)=1+\epsilon \; z_1(\tau) +\cdots $,
choosing $ z_1 $ to cancel the secular term at this time scale leads to
the renormalization 
group equation 
$$
\frac{\partial A(\tau)}{\partial \tau}+\frac{\epsilon}{2} \;A(\tau)=0
$$
and the improved solution $y(t)=
e^{-\frac{\epsilon}{2}t}(A(0)e^{it}+\mbox{c.c})$ after setting
$\tau=t$ in the 
solution. This obviously is the correct solution to ${\cal O}(\epsilon)$.
The interpretation of the renormalization group resummation is very clear
in this simple example: the perturbative expansion is carried out to a time
scale $\tau <<1/\epsilon$ within which perturbation theory is
valid. The correction is recognized as a change in the 
amplitude, so at this time scale the correction is absorbed in a renormalization of the amplitude and the
perturbative expansion is carried out to a longer time but in terms of
the {\em amplitude at the renormalization scale}. The dynamical
renormalization group equation is the differential form of this
procedure of evolving in time, absorbing the corrections into the
amplitude (and phases) and continuing the evolution in terms of the
renormalized amplitudes and phases. This is the same spirit as the
momentum-shell renormalization in critical phenomena. The details
of the second order calculation and implementation of the
renormalization group for this simple problem are offered in appendix
III to illustrate the shift in the frequency.

 This interpretation in the simple exercise extends to the more
 complex situations with the same underlying mechanism, the secular
 divergences are absorbed in the complex amplitudes and the
 perturbative expansion is then carried in terms of the renormalized
 amplitudes. The differential form of this process is the dynamical
 renormalization group equation. Thus the similarities with the usual
 renormalization procedure are manifest.  

In the same manner that the usual renormalization group sums the
leading logarithms when the renormalization group functions are
computed to one loop order, the dynamical renormalization group sums
the leading secular terms when the coefficients are computed to lowest
order. This is manifestly revealed by the naive perturbative expansion
of the real time propagator (\ref{asymptotic}).  

The main objective of studying the different cases in the scalar theory
of the previous section was to thoroughly {\em test} the method with a
non-trivial example that can be studied with different methods. By
studying 
these examples in detail we have learned that the dynamical
renormalization group  provides: i) a real-time equivalent of the
resummation via the renormalization group in Euclidean space time
(equivalent to Bloch-Nordsieck resummation) in the case of infrared
threshold divergences at 
$T=0$ leading to relaxation with anomalous exponents, ii) the usual
mass shift and damping rates in the case of narrow resonances and
iii) leads to 
a similar resummation scheme at $T \neq 0$ when the threshold infrared
divergences are more severe. This detailed analysis then provides
confidence on this novel method to study the more interesting and
relevant case of 
a gauge theory.


\section{A gauge theory: SQED} 
 
We are now in position to apply the method of the dynamical renormalization
group to implement the resummation of infrared divergences in gauge theories, which is our primary goal. 

We will study the case of scalar QED, since to lowest order in  hard
thermal loops this theory has the {\em same} properties as those of
QED and QCD\cite{lebellac,blaizot1,rebhan,iancu}, in particular the
infrared divergences associated with the propagation of the charged fields. 

Since we are primarily interested in studying the real-time manifestation
of the finite temperature infrared divergences, we will focus on the
relaxation of the charged scalar field at finite
temperature. Furthermore we will only consider the contribution of
transverse photons to the charged scalar self energy, since 
 longitudinal photons are Debye screened at finite temperature ($m_D \approx
e T$) and do not contribute to the infrared divergences.
 
In this Abelian theory it is rather straightforward to implement a 
gauge invariant formulation by projecting the Hilbert space on states
annihilated by Gauss' law. Gauge invariant operators can be constructed
and the Hamiltonian and Lagrangian can be written in terms of these. The
resulting Lagrangian is exactly the same as that in Coulomb
gauge\cite{boyhtl} and is given by 
\begin{eqnarray}  
{\cal L}=&&\partial_\mu\Phi^\dagger\,\partial^\mu\Phi -m^2 \Phi^{\dagger}\Phi  
+\frac{1}{2}\partial_\mu \vec{A}_T\cdot\partial^\mu\vec{A}_T 
-e\vec{A}_T\cdot\vec{j}_T 
-e^2\vec{A}_T\cdot\vec{A}_T\; \Phi^\dagger\Phi +\nonumber \\ 
&&+\frac{1}{2}\left(\nabla A_0 \right)^2+ {e^2}A^2_0 \; \Phi^{\dagger}\Phi+ 
eA_0\;\rho\; , \nonumber \\  
\vec{j}_T=&&i(\Phi^\dagger\vec{\nabla}_T\Phi-\vec{\nabla}_T\Phi^\dagger\Phi)
\quad ; \quad \rho = -i\left(\Phi
\dot{\Phi}^{\dagger}-{\Phi}^{\dagger}\dot{\Phi}\right)\;.  \nonumber 
\end{eqnarray} 
where we have traded the instantaneous Coulomb interaction for a gauge invariant Lagrange
multiplier field $A_0$ which should not be confused with a time component
of the gauge field. $\vec{A}_T$ is the transverse component satisfying
$\vec{\nabla}\cdot \vec{A}_T(\vec x,t)=0$. 
Since we are only interested in obtaining the infrared behavior arising
from finite temperature effects we do not introduce the renormalization
counterterms to facilitate the study. The finite temperature behavior is
ultraviolet finite. The non-equilibrium generating
functional requires the fields on the forward and backward branches\cite{boyhtl}. The equation of motion for the
charged scalar field is obtained by writing\cite{tadpole1,boyhtl}
$$
\Phi^{\pm}(\vec x,t) = \varphi(\vec x,t) + \Delta^{\pm}(\vec x,t) \; \;  ; \; \;  \langle \Delta^{\pm}(\vec x,t) 
\rangle = 0 
$$
and similarly for the hermitian congujate fields. 

In obtaining the equation of motion to one-loop order, we neglect the
contribution from the Coulomb interaction. The reason as explained above is that the long-range Coulomb interaction
 will be screened by finite temperature
effects with a Debye screening length $m_D \approx eT$ and hence the screened Coulomb interaction will be free of 
infrared divergences. However
in an Abelian plasma the magnetic (transverse) photons are not screened
(no static screening, only dynamical screening through Landau damping) and
the exchange of soft magnetic photons will lead to threshold infrared
divergences. 

In terms of the spatial Fourier transform of $\varphi(\vec x,t)$ we
find the equation of motion
\begin{eqnarray}
&& \ddot{\varphi}(-\vec k,t)+ (k^2+m^2+e^2 \langle \vec{A}^2_T(\vec
x,t) \rangle ){\varphi}(-\vec k,t)-4ie^2 \int_{-\infty}^{\infty} dt'
\int\frac{d^3q}{(2\pi)^3} \; k^i_T(\vec q) \; k^j_T(\vec q)\times \left\{
\right. \nonumber \\ 
&& \left. \left[\langle A^{+i}_T(\vec q,t) A^{+j}_T(-\vec q,t')\rangle \langle 
\Delta^+(-\vec k-\vec q,t) \Delta^{\dagger,+}(\vec k+\vec q,t')
\rangle - \right. \right. \nonumber  \\ 
&& \left. \left. \langle A^{+i}_T(\vec q,t) A^{-j}_T(-\vec
q,t')\rangle \langle  
\Delta^+(-\vec k-\vec q,t) \Delta^{\dagger,-}(\vec k+\vec q,t') \rangle
\right] \varphi(-\vec k,t') \right\} = J(-\vec k,t) \nonumber
\\ && \vec{k}_T(\vec q) = \vec k - \hat{q}(\vec k \cdot \hat{q})\nonumber
\end{eqnarray}

where we coupled an external source 
$J(\vec k,t) = J(\vec k)e^{\epsilon t} \Theta(-t)$ ($\epsilon \rightarrow 0^+$) to provide an initial value problem 
with an adiabatic switching on of the expectation value. The initial conditions for the expectation value are
$$
\varphi(\vec k,t=0) = \varphi_k(0) \; \; ; \; \; \dot{\varphi}(\vec
k,t=0) =0 
$$

Infrared phenomena is associated with the soft limit of the
intermediate photon, and therefore requires the HTL resummation of the
intermediate photon propagator. However,  we begin our study of SQED by implementing the dynamical renormalization
group to resum the perturbative expansion in the case in which the
self-energy of the charged field only includes the exchange of a bare
transverse photon. The study of this situation will shed light on the
different physical processes that contribute and those that do not because
of dynamical screening via Landau damping. 
 In the next section we will include the HTL resummation
of the exchanged photon and implement the dynamical renormalization group.

\subsection{\bf Bare photon propagators}

The necessary free-field non-equilibrium Green's functions are given by

$\bullet\text{Scalar Propagators}$ 
$$
{\langle}{\Phi}^{(a)\dagger}(\vec{x},t){\Phi}^{(b)}(\vec{x}, 
t^{\prime}){\rangle}=-i\int {d^3k\over{(2\pi)^3}}\; G_k^{ab}(t,t^\prime) \;
e^{-i\vec{k}\cdot(\vec{x}-\vec{x^\prime})}\;, 
$$
where $(a,b)\;\in\{+,-\}$. 
\begin{eqnarray} 
&&G_k^{++}(t,t^\prime)=G_k^{>}(t,t^{\prime})\Theta(t-t^{\prime}) 
+G_k^{<}(t,t^{\prime})\Theta(t^{\prime}-t)\; , \label{gplpl} \nonumber
\\ 
&&G_k^{--}(t,t^\prime)= G_k^{>}(t,t^{\prime})\Theta(t^{\prime}-t)+ 
G_k^{<}(t,t^{\prime})\Theta(t-t^{\prime})\;, \nonumber
\\ 
&&G_k^{\pm\mp}(t,t^\prime)=-G_k^{<(>)}(t,t^{\prime})\;, 
\label{gplmin}\nonumber \\ 
&&G_k^{>}(t,t^{\prime})=\frac{i}{2\omega_k}\left[ 
(1+n_k)\;e^{-i\omega_k(t-t^\prime)} 
+n_k\;e^{i\omega_k(t-t^\prime)}\right],\label{greater} \nonumber
\\ 
&&G_k^{<}(t,t^{\prime})=\frac{i}{2\omega_k}\left[ 
n_k\;e^{-i\omega_k(t-t^\prime)} 
+(1+n_k)\;e^{i\omega_k(t-t^\prime)}\right]\;,\label{lesser} \nonumber
\\ 
&&\omega_k=\sqrt{\vec{k}^2+m^2}\quad;\;\;\;\;\;\;n_k= 
\frac{1}{e^{\beta\omega_k}-1}\;.  \nonumber
\end{eqnarray} 
$\bullet\text{Photon Propagators}$ 
$$
{\langle}{A}^{(a)}_{Ti}(\vec{x},t){A}^{(b)}_{Tj}(\vec{x}, 
t^{\prime}){\rangle}=-i\int {d^3k\over{(2\pi)^3}}\;{\cal G}_{ij}^{ab} 
(k;t,t^\prime)\;e^{-i\vec{k}\cdot(\vec{x}-\vec{x^\prime})}\;, 
$$
\begin{eqnarray} 
&&{\cal G}_{ij}^{++}(k;t,t^\prime)={\cal P}_{ij}(\vec{k}) \;
\left[{\cal G}_k^{>}(t,t^{\prime})\Theta(t-t^{\prime}) 
+{\cal G}_k^{<}(t,t^{\prime})\Theta(t^{\prime}-t) \right]\;,
\label{phot++}\\  
&&{\cal G}_{ij}^{--}(k;t,t^\prime)= {\cal P}_{ij}(\vec{k}) \;
\left[{\cal G}_k^{>}(t,t^{\prime})\Theta(t^{\prime}-t) 
+{\cal G}_k^{<}(t,t^{\prime})\Theta(t-t^{\prime}) \right]\;, 
\label{phot--}\nonumber\\ 
&&{\cal G}_{ij}^{\pm\mp}(k;t,t^\prime)=-{\cal P}_{ij}(\vec{k}) \;
{\cal G}_k^{<(>)}(t,t^{\prime})\;, \label{photpm}\\ 
&&{\cal G}_k^{>}(t,t^{\prime})=\frac{i}{2k}\left[ (1+N_k)e^{-ik(t-t^\prime)} 
+N_ke^{ik(t-t^\prime)}\right]\;,\label{phot>} \nonumber
\\ 
&&{\cal G}_k^{<}(t,t^{\prime})=\frac{i}{2k}\left[ 
N_k\;e^{-ik(t-t^\prime)} 
+(1+N_k)\;e^{ik(t-t^\prime)}\right],\label{phot<}\nonumber\\ 
&&N_k=\frac{1}{e^{\beta k}-1}\;. \nonumber
\end{eqnarray} 
Here ${\cal P}_{ij}(\vec{k})$ is the transverse projection operator: 
$$
{\cal P}_{ij}(\vec{k})=\delta_{ij}-\frac{k_ik_j}{k^2}\;.
$$

Finally we find the equation of motion for $ t>0 $ to be given by
\begin{eqnarray}
&& \ddot{\varphi}(-\vec k,t)+ [k^2+M^2(T)] \; {\varphi}(-\vec
k,t)+\int_{-\infty}^{t} \Sigma(\vec k,t-t') \; {\varphi}(-\vec k,t') \;
dt' = 0 \label{eqnsqed}\nonumber  \\ 
&& \Sigma(\vec k,t-t')= -8e^2 k^2 \int\frac{d^3q}{(2\pi)^3}
\frac{1-\cos^2 \theta }{4 \; q \; \omega_{\vec k+\vec q}}\left\{ 
\left(1+N_q+n_{\vec k+\vec q}\right)\sin(\omega_{\vec k+\vec
q}+q)(t-t') \right. \nonumber \\ 
&& \left. +\left(N_q - n_{\vec k+\vec q}\right)\sin(\omega_{\vec
k+\vec q}-q)(t-t')  \right\} \label{selfsqed}\nonumber \\
&& M^2(T) = m^2 + e^2 < \vec{A}^2 > \nonumber
\end{eqnarray} 
and $\theta$ is the angle between $\vec k$ and $\vec q$. We see that
the self-energy has a form very similar to that of the scalar case, eq.
(\ref{1lupselfen}), the only difference being the $k^2$ in front (reflecting
the exchange of transverse photons)  and
the $ 1-\cos^2 \theta $ inside the integral. Integrating by parts and
using the initial conditions the resulting equation of motion can be written
in the same form as in eq. (\ref{fineq}) with $\gamma_k(t)$ as in eq. (\ref{gamma}) in terms of the spectral density 
\begin{eqnarray}
\rho(k;\omega)   =    
-\frac{e^2 k^2}{2\pi^2} \int \frac{q
dq}{\omega_{k+q}}\;  (1-\cos^2\theta) \; d\cos\theta\;  & &  
 \left\{
(1+N_q+n_{k+q}) \;\delta\left(\omega-q-\omega_{k+q}\right)
\right. \nonumber \\ 
&- &  \left. 
 (N_q-n_{k+q}) \;\delta\left(\omega-q+\omega_{k+q}\right)\right\}
\label{rhosqed} 
\end{eqnarray}

As in the scalar case, we are mainly interested in the infrared effects associated with the emission and absorption of soft photons in the intermediate state and only the finite temperature contribution. Therefore we will: i) neglect the contribution of the 
distribution function of the intermediate charged scalar, i.e. the term $n_{k+q}$, ii) replace $N_q \approx \frac{T}{q}$ in the expression and neglect the vacuum contribution
(the one), thus obtaining
$$
 \rho(k;\omega)= 
-\frac{e^2 kT}{2\pi^2} \left\{ \int^{q^+}_{q^-} \frac{dq}{q} \;
[1-X^2(\omega,q)] \; \Theta(\omega-\omega_k)- \int_{q^+}^{q^c} 
\frac{dq}{q} \; [1-X^2(\omega,q)] \;\Theta(k^2-\omega^2) \right\}
$$
with $q^c$ an upper momentum cutoff $q^c <<T$ and 
\begin{eqnarray}
 X(\omega,q) = && \frac{\omega^2-\omega^2_k-2\omega q}{2kq}
\quad , \nonumber \\
 q^{\pm}  = && \frac{\omega^2 - \omega^2_k}{2(\omega\mp k)}\quad . \nonumber
\end{eqnarray}
The second contribution with support below the light cone is identified
with the Landau damping cut. Since the time dependent correlation functions
involve the product of $\rho(k;\omega) \cos\omega t $ the frequency
integral in the range $ -k < \omega <0 $ arising  
from the Landau damping cut can be combined with the contribution from
the positive frequency range by the symmetry of the integrand in eq. (\ref{fi1atime}). After a straightforward 
calculation we finally find the
finite temperature contribution to the spectral density near threshold
to be given by
\begin{equation}
\rho_{I.R.}(k;\omega) = -\frac{e^2}{2\pi^2}\left(\frac{T}{k}\right)
\left\{  (\omega^2 - k^2)\ln\left|\frac{\omega-k}{\omega+k}\right|
+ 2 k\omega\right\}
\left[\Theta(\omega-\omega_k)+\Theta(k-\omega)\right]\Theta(\omega)
\label{specdenssqed} 
\end{equation}

There are several noteworthy features of these spectral density near 
threshold, as compared to the simpler case of the scalar theory studied
in section III (see eq. (\ref{effspecscal}). In this case the spectral
density is constant at threshold. The term $\omega^2$ (multipling the
logarithms in (\ref{specdenssqed}))  and the last term $8 \omega k$
arise from the region of the integral for which $ \cos^2\theta
\approx 1 $ which 
would give a vanishing integrand were it not for the fact that there is
a linear divergence as $\omega \rightarrow k$. These contributions arise
from the emission and absorption of photons which are almost collinear with
the incoming charged scalar.  

We will now  study the case of a massive scalar in two  important
 limits: i) $k/\omega_k =v << 1$ ($m \neq 0$),  
 ii) $k/\omega_k =v \approx 1$.  In
both cases the leading contribution is easily recognized to arise from
the production cut $\omega \geq \omega_k$. For $m \neq 0$ there are no
infrared divergences associated with the Landau damping cut. 

\subsubsection{$v << 1$} 

\begin{figure}[t] 
\epsfig{file=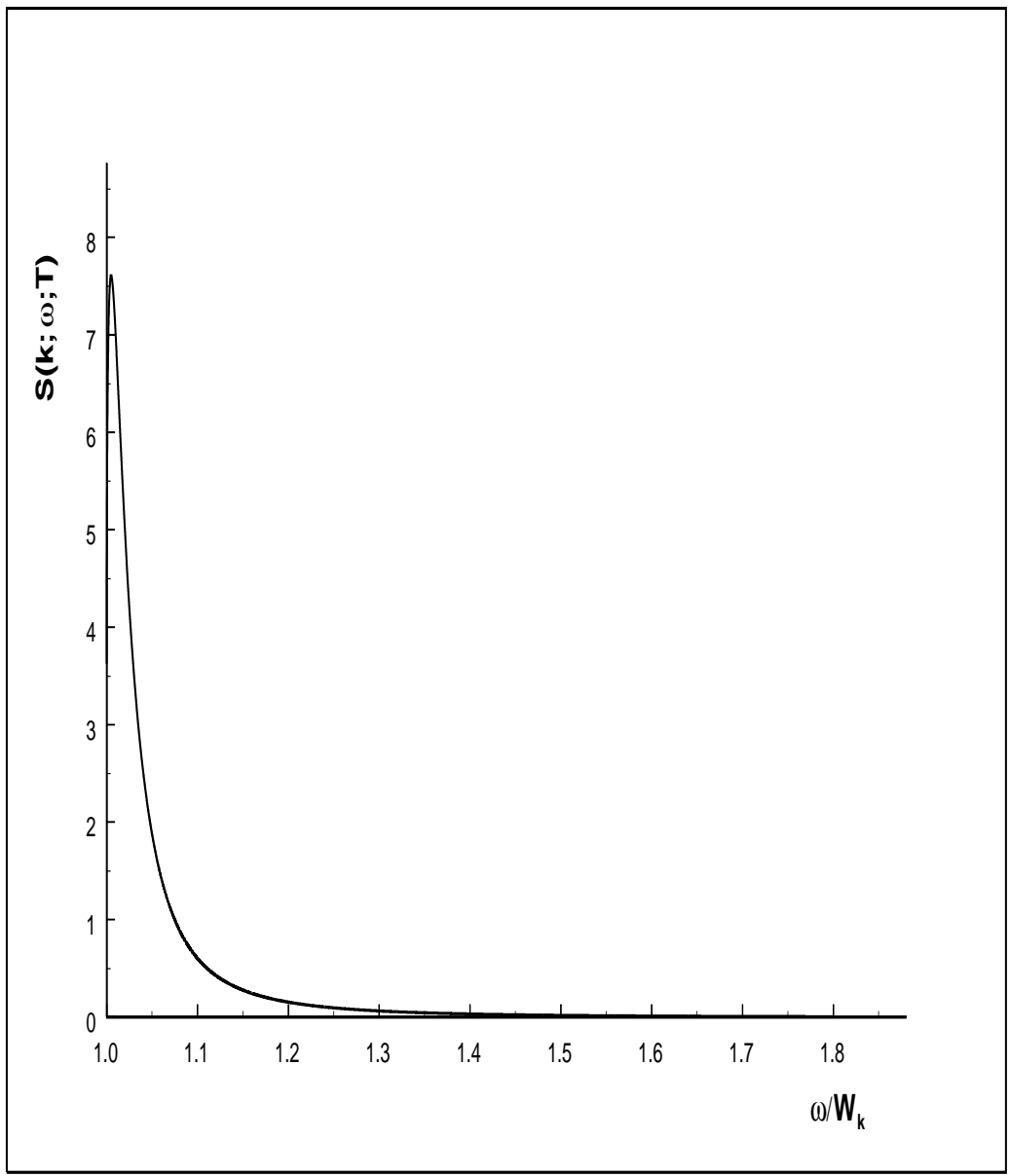,width=15cm,height=18cm} 
\caption{ The spectral density for the $\sigma$ field $S(k,\omega;T)$
(eq. (\ref{sigmaspec}))  near threshold for $\rho(k,\omega_k) =
-0.005$. \label{fig1}} 
\end{figure}

 Since the spectral
density is slowly varying near threshold we can find the asymptotic 
behavior  in time  of the coefficient of $A_k \, e^{i\omega_k t}$ in
(\ref{fi1atime})  at large times following the same method as in 
previous sections and appendix B. We find the following results:
i) the imaginary part  of the coefficient of $ A_k \; e^{i\omega_k t} $
is given by 
\begin{eqnarray} \label{imagsqed} 
&&\frac{i\;  t}{4} \int_{\omega_k}^{\infty} d\omega \;
\frac{\rho(k;\omega)}{\omega} \frac{1}{\omega-\omega_k}\left[1- 
\frac{\sin(\omega-\omega_k)t}{(\omega-\omega_k)t}\right]
\buildrel{\mu t >> 1}\over=
-\frac{i\, e^2\, T}{8\pi^2} f(v) \; t \ln\overline{\mu}t\cr \cr
&&+ i\; t \int_{\omega_k}^{\infty} {d\omega  \over \omega - \omega_k}\left
[ \frac{\rho(k;\omega)}{\omega} -
\frac{\rho(k;\omega_k)}{\omega_k}\Theta(\mu-\omega+\omega_k) \right]
+ {\cal O}\left( {e^2 \, T \over t } \right)
\end{eqnarray}
with
\begin{eqnarray}
f(v) & = & \frac{1-v^2}{v}\ln\frac{1-v}{1+v} \; +2 \; \; ; \;
\; v=\frac{k}{\omega_k} \nonumber \\ 
\overline{\mu} & = & \mu\; e^{\gamma -1}
\end{eqnarray}
\noindent the function $0< f(v) <2$ for  $0< v \leq 1$.  

ii) The real part of the coefficient of $ A_k \; e^{i\omega_k t} $ in the
asymptotically large time limit is given by (see appendix B)
\begin{equation}
 \frac{1}{4} \int_{\omega_k}^{\infty} \frac{d\omega}{\omega} \; \rho(k;\omega)
\; \frac{1-\cos(\omega-\omega_k)t}{(\omega-\omega_k)^2}\buildrel{\mu t >>
1}\over= -\Gamma_k \; t + {2\, \over \pi \,\omega_k}\log(\mu \; t \;
e^{\gamma}) \left[ \Gamma_k - \frac14 {\partial
\Sigma_I\over\partial\omega}(k, \omega_k)\right] + {\cal O}\left( e^2
\, T \right) 
\label{realsqed}
\end{equation}
with $ \Gamma_k $ given by
\begin{equation}
\Gamma_k= -\frac{\pi}{4} \frac{\rho(k;\omega_k)}{2\omega_k} =
\frac{1}{2}\frac{\Sigma_I(\omega_k)}{2\omega_k}= 
\frac{\pi}{8} \left(\frac{e^2T}{2\pi^2}\right)
f(v) \label{gammasqed}
\end{equation}
where we used the expression (\ref{sigmaimag}) and the fact that for the
production cut, the spectral density only has support for positive frequencies.
We then find the same phenomenon as in the previous case of the scalar
particles in that the damping rate is 
one half of the expected value. The reason again is that the full
spectral density for the charged scalar is very 
similar to that featured in figure (\ref{fig1}) with a prominent peak
near threshold that is almost half of  
a Breit-Wigner peak. However the logarithmic phase clearly exhibits
the fact that cannot be interpreted as a quasiparticle resonance. 

The contributions from the coefficient of $A^*_k e^{i\omega_k t}$ and
the linear secular term from $\phi^{(1,b)}_k$ that 
contributes to the imaginary part are both subleading. We now
renormalize the amplitude as in eqs. (\ref{amplitude})-(\ref{zetadyn})
with the choice 
\begin{eqnarray}
\lambda \; z^{(1)}_I(\tau) & = & \frac{e^2T}{8\pi^2} \; f(v)  \; \tau
\ln\overline{\mu}\tau 
\nonumber \\ \lambda \; z^{(1)}_R(\tau) & = &  \Gamma_k \; \tau  -{2\,
 \over \pi 
 \,\omega_k}\log(\mu \; \tau \; e^{\gamma}) \left[ \Gamma_k -
 \frac14 {\partial  \Sigma_I\over\partial\omega}(k, \omega_k)\right]
\nonumber 
\end{eqnarray}
The solution of the dynamical renormalization group equation now leads
to the asymptotic behavior of the expectation value of the charged fields

\begin{eqnarray}
\phi_k(t) & = &  {\cal A}_k(t_0) \; e^{i\varphi_k(t,t_0)} \; 
e^{-\Gamma_k (t-t_0)} \; \left({t_0 \over t} \right)^{b_k}+ \mbox{c.c.} \; ,
\nonumber \\
\varphi_k(t,t_0) & = & \omega_{k,R} \; t - \frac{e^2T}{8\pi^2} \;  f(v) \; t
\; \ln\frac{t}{t_0} \; ,
\cr b_k &=&{2\, \over \pi \,\omega_k}\left[ -\Gamma_k + \frac14 {\partial
\Sigma_I\over\partial\omega}(k, \omega_k)\right]\; .\label{htlsuave}
\end{eqnarray}
Where we integrated the dynamical renormalization group equation with 
initial condition at $t_0$ which is taken as some arbitrary renormalization
point replacing the infrared cutoff $\overline{\mu}$. The renormalization
group invariance of $\phi_k(t)$ is now explicit, a change of the arbitrary
scale $t_0$ is compensated by a change in the amplitude ${\cal A}_k(t_0)$.

\subsubsection{$v \rightarrow 1$} 
The limit $v \rightarrow 1$ must be studied carefully.
As $\omega \rightarrow \omega_k \approx k$ the term $ (\omega^2 -
k^2)\ln|\omega-k| $ cannot be taken outside of the integral. However
upon  
the change of variable $ \omega -k = z / t $ in the integral in the same
manner as that leading to (\ref{realsqed}) leads to a term that is
of the form $ \ln[\mu t]/t$ for this contribution, which then becomes
subleading compared with the term in $\rho_{I.R.}(k;\omega)$ that does
not vanish as $v \rightarrow 1$. The asymptotic large time behavior is
therefore obtained from the previous section with $v\neq 1$ by simply
setting $v=1$. The contribution to the final result in this limit
arises solely from the emission and absorption of collinear photons.

It is illuminating to try and understand this result in the hard
limit $v\rightarrow 1$. The delta functions in (\ref{rhosqed}) in the
limit $k \gg m$ become $\delta(\omega -k -q\mp q \cos\theta)$ therefore
as $\omega \rightarrow k$ the whole contribution arises from photons
that are emitted or absorbed collinearly $\theta = 0, \pi$ with the
moving (hard) scalar. However, as pointed out originally by Pisarski\cite{robinfra},
the contribution from collinear photons does not survive screening effects
arising from higher order contributions to the photon propagator. In particular, dynamical screening as a consequence of Landau damping of the
intermediate photons cuts off the contribution of collinear photons, and
lead to a greater contribution of photons emitted or absorbed at right
angles with respect to the moving charged particle. 

The analysis of this section is illustrative of the power of the dynamical renormalization group to obtain
the asymptotic long time behavior. For the case of QED, QCD or SQED, the analysis presented in this section
in terms of the bare propagators for the scalars and photons has very limited validity. For soft external momentum,
the infrared region of the internal loop requires HTL resummation of the internal lines and vertices\cite{robinfra,iancu,rebhan}.
The main purpose of our analysis in this section, however,  was to illustrate how the dynamical renormalization group is capable
of revealing novel forms of relaxation with logarithmic corrections, power laws, anomalous dimensions etc. These alternative
forms of relaxation cannot be found by attempting to describe exponential relaxation and computing an imaginary part of
the self-energy on shell. 

We now include screening corrections via HTL resummation of internal lines. We will consider the case of hard external momentum for which only the internal photon line must be HTL resummed (the scalar is massive and hard).

\subsection{\bf Hard thermal loop-resummed photon propagators}

We now focus on the case of hard external momentum of the charged scalar. In this case only the internal photon
line receives HTL corrections, since the scalar in the loop is massive and hard, the vertex does not require
resummation because one of the momenta into the vertex is hard (the scalar)\cite{robinfra,iancu,rebhan}. Hence this situation is simpler than
the case of soft external momentum that will be studied elsewhere. 

In order to include
the leading order screening effects in the photon propagator, we must use
the hard thermal loop resummed propagators\cite{robinfra}. 
The generalization of the HTL resummation program in the Matsubara
formulation of finite temperature field theory to the real time 
 formulation is described in detail in  appendix A, we  collect
here only the main ingredients.

The photon propagators can be written as in (\ref{phot++})-(\ref{photpm})
but now with the resummed Wightmann functions (see appendix)
\begin{eqnarray}
{\cal G}_q^>(t-t') & = & \int dq_0 \;  \tilde{\rho}_T(q_0,q)\;\left[ 1
+N(q_0) \right] \;
e^{-iq_0(t-t')} \nonumber \\
 {\cal G}_q^<(t-t')& = & \int dq_0 \; \tilde{\rho}_T(q_0,q)\;N(q_0) \;
e^{-iq_0(t-t')} \nonumber 
\end{eqnarray}
where in the hard thermal loop limit the
spectral density for  transverse photons is given by\cite{rebhan,boyhtl}

\begin{eqnarray}
\tilde{\rho}_T(q_0,q) & = &
\frac{1}{\pi}\frac{\Sigma_I(q_0,q)\;\Theta(q^2-q^2_0)}{\left[q^2_0-q^2
-\Sigma_R(q_0,q)\right]^2+\Sigma^2_I(q_0,q)} +\mbox{sign}(q_0)\;
Z(q)\;\delta(q^2_0-\omega^2_p(q)) \label{htlrho} \cr \cr
\Sigma_I(q_0,q) & = & \frac{\pi e^2
T^2}{12}\frac{q_0}{q}\left(1-\frac{q^2_0}{q^2}\right) 
\cr \cr 
\Sigma_R(q_0,q) & = & \frac{ e^2
T^2}{12}\left[2\frac{q^2_0}{q^2}+\frac{q_0}{q}
\left(1-\frac{q^2_0}{q^2}\right)
\ln\left|\frac{q_0+q}{q_0-q}\right|\right]\nonumber
\end{eqnarray}
where $\omega_p(q)$ is the plasmon pole and $Z(q)$ its (momentum dependent)
residue, which will not be relevant for the following discussion. 
Inserting these propagators in the expression for the self energy, 
keeping only the term $N(q_0) \approx T/q_0$ in the resulting expressions, and
focusing on the hard scalar limit $k \approx T \gg m$ we find that the
self energy can be written in the form of a dispersion relation just
as in eq. (\ref{sigmadisp}) but with
$$
\rho(k;\omega) = -\frac{e^2 T k}{\pi^2} \int q^2 dq \; \int^1_{-1}dX\;
(1-X^2)
\int dq_0\; \frac{\tilde{\rho}_T(q_0,q)}{q_0}\; \delta(\omega -k - q_0
-qX)  
$$

The infrared region corresponds to small $q$, and we find that for $q << eT$ the integrand $\tilde{\rho}_T(q_0,q)/q_0$ is strongly peaked at $q_0 =0$.
Fig. 2 shows $ \tilde{\rho}_T(q_0,q) q^2/q_0  $ vs. $ q_0 $ in units of
$eT/\sqrt{12}$. We find that for $q << eT$ the photon spectral density
is well approximated by\cite{iancu}
\begin{eqnarray}
\frac{\tilde{\rho}_T(q_0,q)}{q_0} & = & \frac{1}{\pi q^2}
\frac{\Gamma}{\Gamma^2+q^2_0} \nonumber \\
\Gamma & = & q^3 \frac{12}{\pi e^2 T^2} \nonumber
\end{eqnarray}
When this  Lorentzian distribution is integrated with smooth functions, it
can be expanded in  the width  obtaining
$$
 \frac{1}{\pi}\frac{\Gamma}{\Gamma^2+q^2_0} \approx  \delta(q_0) -
 \frac{\Gamma^2}{2} \delta''(q_0)+\cdots 
$$

\noindent and the infrared behavior is dominated by the $\delta(q_0)$. Finally
we find the infrared behavior of the spectral density of the self-energy
to be given by
\begin{equation}
\rho(k;\omega) \approx -\frac{ e^2 k T}{\pi^2 } \int^{q^*}_{|\omega-k|}
\frac{dq}{q} \left(1- \frac{(\omega-k)^2}{q^2}\right) \approx \frac
{ e^2 k T}{\pi^2} \ln\frac{|\omega-k|}{\mu} 
\label{infrahtlspec} 
\end{equation}
where $q^* \sim \mu \leq eT$ is an arbitrary upper momentum cutoff, which physically is of the order of the
plasma frequency. We thus
see that dynamical screening originating in Landau damping for the photon
propagator suppresses the contribution of collinear photons and as
$\omega \rightarrow k$ the contribution to the self energy arises
primarily from 
photons emitted or absorbed at right angles\cite{robinfra}.
This is the same situation as in QED\cite{iancu}.

 The final spectral density
for the self energy, eq. (\ref{infrahtlspec}) is therefore of the
{\em same form} as the finite temperature self energy of the simple
scalar theory 
studied in the earlier sections, see eq. (\ref{effspecscal}), and
thus justifies our excursion into that simpler theory.

We can now follow the same steps to study the secular terms in the
real time perturbative expansion, which lead to
eq. (\ref{longtime}). 

The long time asymptotic behavior is obtained by inserting the
spectral density eq. (\ref{infrahtlspec}) in the expressions
(\ref{fi1atime})-(\ref{fi1btime}) with $ \omega_k = k $. The asymptotic
dependence on time is extracted by changing variables to $ \omega -k =
z/t $ 
and upon setting $ A_k = A^*_k $, we find: i) the {\em imaginary}
contribution to $\phi^{(1,a)} + \phi^{(1,b)}$ is given  
by an {\em infrared finite} and linear in time secular term which is
interpreted as a renormalization of the frequency.  
ii) the {\em real} contribution to $\phi^{(1,a)} + \phi^{(1,b)}$ is
given by  
$$
\mbox{Re}\, \delta\phi_k(t) = -\alpha\,  T\,  t\, \ln\mu t \; \; ; \; \;
\alpha= \frac{e^2}{4\pi} 
$$
and the first order correction is thus found to be given by
$$
\phi^{(1)}_k(t) = A_k e^{i\omega_k t} \left[i \;  \delta \omega_k \; t -
\alpha \; T \; t \;\ln\mu t \right] + \mbox{c.c} + \mbox{regular
perturbative terms} 
$$
Introducing the renormalization of the amplitude as in equations
(\ref{amplitude})-(\ref{zetadyn}) and choosing 
$$
\lambda z^{(1)}_R(\tau) = \alpha \; T \; \tau \;\ln\mu \tau \; ; \; \; 
\lambda z^{(2)}_R(\tau) = - \delta \omega_k \; \tau 
$$

\noindent we obtain the renormalization group equation

$$
\frac{\partial {\cal A}_k(\tau)}{\partial \tau} - \left[i \; \delta
\omega_k  -\alpha \; T \; (\ln\mu \tau+1)  \right]{\cal A}_k(\tau) =0
$$
with solution
$$
{\cal A}_k(\tau) ={\cal A}_k(t_0)\;  e^{i\delta \omega_k (\tau - t_0)} \; 
e^{-\alpha T \tau \ln\frac{\tau}{t_0}} 
$$
\noindent where we have chosen $ t_0 = \mu^{-1} $ as an initial condition
for the integration. Now choosing the arbitrary renormalization scale
$\tau$ to coincide with the time $t$ we finally arrive at one of the
main results of this article which is the asymptotic behavior in time
of the expectation value of the charged scalar field

\begin{equation}
\phi_k(t)  =   {\cal A}_k(t_0) \;e^{i\tilde{\omega}_k(t-t_0)} \;
e^{-\alpha T t\ln[t/t_0]} + \mbox{c.c.}
\label{asihtl}
\end{equation}
 thus displaying the renormalization group invariance of
the solution: a change of the arbitrary time scale $t_0$ (inverse of the
infrared cutoff) is compensated
by a change in the amplitude. The quantity $\tilde{\omega}_k$ is the
renormalized mass including the (infrared finite) HTL corrections. 

A similar behavior for the asymptotic dynamics of the {\em fermion} field
in QED 
has been obtained via the Bloch-Nordsieck resummation in reference\cite{iancu}. The power of the dynamical 
renormalization group
has now become explicit in that it transcends any approximation and implements a resummation of the logarithmic 
infrared divergences that
in this case includes the resummation of the hard thermal loops. 

\section{Conclusions, comments and more questions:}

In this article we have introduced a novel method 
of dynamical renormalization group resummation to study relaxation
in real time. The first step of the program is to relate the retarded
Green's functions that contain the dynamical information on the
time evolution away from equilibrium in linear response to an initial value problem for
the {\em expectation value} of the fields. This initial value problem 
in real time allows to implement the method of dynamical renormalization
 improving  the perturbative solution by  a resummation. This resummation
is the real time counterpart of the resummation via the renormalization
group in Euclidean field theory. We first apply our methods to 
a scalar theory of one massive and one massless field. The emission and
absorption of the massless field introduced infrared divergences akin
to those found in gauge theories. We compare in such a model the late
behaviour of the massive field amplitude at zero temperature
[eq.(\ref{ampliesc})]  
to results  based on Bloch-Nordsieck resummations [eq.(\ref{asymptotic})]
as well as the renormalization group applied to Euclidean Green's
functions [eq.(\ref{resu2p})]. 

Furthermore, for large temperature we compute the the late
behaviour of the massive field amplitude both for  hard and soft 
external momenta. For hard modes, the field amplitude relaxes as $
e^{-(g^2  \, T /4\pi ) \, t  \,\ln[t/t_0]} $ where $ g $ stands for the
coupling constant.  

In real time the infrared divergences are manifest as secular terms in the
perturbative solution of the evolution equation for the expectation value
of the fields. The dynamical renormalization group implements a resummation
of this secular terms that leads to an asymptotically convergent
solution and clearly describes relaxation in real time.  

After applying our method to the scalar model, we focused our attention
to implementing the dynamical renormalization group to resum the infrared
divergences associated with massless transverse photons in scalar QED at
finite temperature. The infrared divergences in this theory are similar
to those found in QED and in lowest order in QCD. 

We have included the resummation of the hard thermal loops and Landau
damping in the internal transverse photon propagators, and implemented
a dynamical renormalization group resummation. The renormalization group
improvement leads to an anomalous
logarithmic relaxation for hard modes as a consequence of infrared
divergences associated 
with the emission and absorption of photons at right angles. 
These anomalous logarithmic relaxation are similar to the scalar field
behaviour and consistent with those found in QED
via the Bloch-Nordsieck resummation\cite{iancu}.

In all cases investigated (both in the scalar model and in QED) the
field behaviour prevents an
interpretation of the relaxation of charged excitations in the medium in the form of a simple 
exponential with  a damping rate determined by the imaginary
part of the self-energy on-shell.  

The advantage of the dynamical renormalization group is that its
implementation is rather simple and trascends any approximations of
the Bloch-Nordsieck type, it can be consistently improved by considering higher
orders in the hierarchy of equations obtained in perturbation theory. 

Furthermore, the real time dynamics obtained via this resummation program
leads to a clear interpretation of the relaxational processes and time
scales {\em without} any assumptions on the validity of the quasiparticle
picture of collective excitations. 
The analysis of secular terms in lowest order provides a simple criterion
for deciding if the collective excitations can be described as narrow
resonances with a width determined by the imaginary part of the
self-energy on shell: linear secular terms lead to such a
quasiparticle description, non-linear secular terms in lowest order
signal anomalous, non-exponential relaxation. 

The dynamical renormalization is 
a different resummation scheme than the HTL resummation, and the latter
can be consistently included in the former as was shown in this article.

We are currently implementing this method to study relaxation of soft and hard fermion
and gauge fields in QED and QCD directly in real-time, thus bypassing the
conceptual limitations of the quasiparticle picture in the sense of exponential relaxation and a
 damping rate determined by the imaginary part of the self-energy on-shell. We expect to report
results in the near future. 


{\bf Comments and further questions:} As we have seen in the example worked out in detail in section III, at finite temperature the infrared divergences are akin to those of a {\em superrenormalizable} theory of critical phenomena, in the sense that
they are no longer logarithmic because the temperature introduces a new scale. There are very few methods for renormalization of {\em infrared} divergences in superrenormalizable theories near the critical point, one of the most
popular being the $\epsilon$ expansion where $\epsilon=d-4$ is the departure from the upper critical dimension. Whereas
the validity of the epsilon expansion appended by Pade resummation has
been confirmed in Ising-like models via either strong coupling lattice
expansion or Montecarlo simulations, the validity in a general case is
at best questionable for $\epsilon=1$. The $\epsilon$ expansion resums
some subset of the Feynman diagrams and only some part of
them\cite{amit} the leading logarithms. 

The dynamical renormalization group therefore provides an alternative
to study the infrared divergences directly in real time by resumming
secular terms in the perturbative solution of the equation of
motion. In the case of infrared divergences the secular terms reflect
these in the form of logarithmic dependence on time. However, the
uselfullness of the dynamical renormalization group is not restricted
to this logarithmic divergences, as explicitly shown in section III,
in the usual case of narrow resonances, the secular terms are linear
(in lowest order) and their resummation through the renormalization
group equation leads to the usual quasiparticle real-time evolution.

There is a translation between the resummation  implied by the usual
(Euclidean) renormalization group and that by the dynamical version: 
the Euclidean version sums the leading logarithms\cite{wein,amit} the
dynamical version sums the leading secular terms\cite{goldenfeld}.

It has been shown in ref.\cite{goldenfeld} using the formal theory of envelopes that the dynamical
renormalization group resummation of secular terms provides an uniform
approximation to the exact solution for systems of ordinary
differential equations. This is true to any given order 
of perturbation for arbitrary ordinary differential equations\cite{goldenfeld}.
 It will be very interesting to extend such a proof to
the evolution equations considered in the present paper. 
 
Thus, for the moment,  the situation with the dynamical RG is similar to that of the
$ \epsilon $ expansion in critical phenomena for $ \epsilon=1 $: it
provides a resummation scheme for the infrared behavior in a
consistent manner and it agrees with known results in cases where it can be compared.  Furthermore in the
case in which the renormalization group and Bloch Nordsieck lead to non-trivial exponentiation of infrared
divergences, the dynamical RG reproduces the results in real time. Thus we believe that the cases analyzed in detail 
in this article and those analyzed in the literature provide very strong evidence for the validity of this approach.
The promise of the dynamical RG as a powerful method to study transport phenomena warrants a deeper study on the
renormalization aspects of the evolution equations in real time and a more formal proof of the applicability of the 
dynamical RG in these problems. This avenue of study is currently in progress.  

\section{acknowledgements}
D. B. thanks the N.S.F. for partial support through grants PHY-9605186 and INT-9815064 and 
LPTHE at the Universit\'e Pierre et Marie Curie for hospitality. 
R. H. is supported by DOE grant DE-FG02-91-ER40682. M. Simionato thanks  LPTHE for kind hospitality,
foundation Aldo Gini and INFN for financial support. 
The authors acknowledge support from NATO.

\appendix
\section{\bf Exact retarded propagators}
In this appendix we gather and generalize some results of the HTL resummation programme
in Matsubara finite temperature field theory\cite{lebellac}, to real time. 

Consider the {\em exact} equilibrium Wightman and retarded Green's 
functions for a real scalar field
\begin{eqnarray}
-iG^>_k(t-t') & = & \langle \Phi_{\vec k}(t) \Phi_{-\vec k}(t') \rangle 
= -i\int d\omega \;  \tilde{G}^>_k(\omega) \; e^{-i\omega (t-t')}\nonumber \\
-iG^<_k(t-t') & = & \langle \Phi_{-\vec k}(t') \Phi_{\vec k}(t) \rangle =
-i\int d\omega \;  \tilde{G}^<_k(\omega) \; e^{-i\omega (t-t')}\nonumber \\
-iG_{R,k}(t-t') & = & -i\left[G^>_k(t-t') -  G^<_k(t-t')\right] 
\Theta(t-t')= -i\int \frac{dq_0}{2\pi} \; \tilde{G}_{R,k}(q_0)
\;e^{-iq_0 (t-t')}  \nonumber
\end{eqnarray}

Inserting a complete set of eigenstates of the full interacting Hamiltonian,
we obtain the spectral representations for the Fourier transforms given
by
\begin{eqnarray}
-i\tilde{G}^>_k(\omega) & = & \frac{1}{\cal Z}\sum_{m,n}e^{-\beta E_m}
\left|\langle m |\Phi_k(0) |n\rangle\right|^2 \delta(\omega-(E_n-E_m))
\nonumber \\ 
-i\tilde{G}^<_k(\omega) & = & \frac{1}{\cal Z}\sum_{m,n}e^{-\beta E_n}
\left|\langle m |\Phi_k(0) |n\rangle\right|^2
\delta(\omega-(E_m-E_n))=e^{-\beta \omega} 
\left(-i\tilde{G}^>_k(\omega)\right)  \label{tilgless} 
\end{eqnarray}
\noindent where the last equality, the KMS condition is obtained by relabelling $m \rightarrow n$ in the sum and 
${\cal Z}=\sum_m e^{-\beta E_m}$ is the equilibrium partition function.   Inserting a representation
of the theta function we finally obtain
\begin{eqnarray}
\tilde{G}_{R,k}(q_0) & = & -\int d \omega \frac{\tilde{\rho}(q_0,k)}{q_0-\omega+i\epsilon} \nonumber \\
 \tilde{\rho}(\omega,k)   & = &
 -i\left[\tilde{G}^>_k(\omega)-\tilde{G}^<_k(\omega)\right] =
 \frac{1}{\pi} \mbox{Im}[\tilde{G}_{R,k}(\omega)]\nonumber
\end{eqnarray}
Using the KMS condition (\ref{tilgless}) we can finally write the {\em
exact} non-equilibrium Wightmann functions in terms of the {\em exact}
spectral density as follows
\begin{eqnarray}
\langle \Phi_{\vec k}(t) \Phi_{-\vec k}(t') \rangle  & = & \int dq_0\;   \tilde{\rho}(q_0,k)\; \left[ 1 +N(q_0) \right]\; 
e^{-iq_0(t-t')} \nonumber \\
\langle \Phi_{-\vec k}(t') \Phi_{\vec k}(t) \rangle  & = & \int dq_0\;   \tilde{\rho}(q_0,k)\; N(q_0) \; 
e^{-iq_0(t-t')} \nonumber\\
N(q_0) & = & \frac{1}{e^{\beta q_0}-1} \label{exactrepre}
\end{eqnarray}
Using the KMS condition, and relabelling the sum indices in the spectral
representation we find that $\tilde{\rho}(q_0,k)=-\tilde{\rho}(-q_0,k)$.The same steps lead to an equivalent 
expression for the transverse
gauge fields, whose Wightmann and Green's functions are proportional to
the transverse projection operator. 

The advantage of the representation (\ref{exactrepre}) is that once we
compute the spectral function $\tilde{\rho}(q_0,k)$ in some approximation,
we can insert the Wightmann functions in the internal loops thus providing
a resummation of the perturbative series. 
The results from the first section allow to obtain the retarded propagator $\tilde{G}_{R,k}(q_0)$ from the solution of
 the initial value problem with
an external source through the relation to linear response as detailed
in the first section. For example by studying  the equation of evolution
for the expectation value of the transverse photon fields in the HTL
approximation as was done in\cite{boyhtl} we can obtain the
spectral representation of the transverse fields in the HTL approximation
and the results of this appendix allow us to implement a resummation of screened photon propagators into the real time description. 

\section{\bf Asymptotic behaviour of spectral integrals}

We summarize in this appendix the late time behaviour of integrals
over the density of states used in Sections III and IV.

In the formulas below $ p(y) $ stands for a smooth function for $ 0
\leq y \leq \infty $. $ p(0) $ as well as $ p'(0) $ are finite. For
large $ y \; ,  p(y) $ decreases as a power such that the integrals
over $ y $  converge  at infinity. These properties are fullfilled in
all cases where these formulae were used in the paper.

\begin{eqnarray}
\int_0^{\infty} { dy \over y^2} \left( 1 - \cos yt \right) \; p(y)
&\buildrel{t \to 
\infty}\over=& \frac{\pi}2 \; t \; p(0) + p'(0) \; \left[ \ln(\mu \,
t) + \gamma \right] \cr \cr  
&+& \int_0^{\infty} { dy \over y^2} \left[  p(y) -  p(0)
- y  \; p'(0) \;\theta(\mu - y ) \right] + {\cal O}\left( {1 \over t }
\right) \; ,  \cr \cr  
\int_0^{\infty} { dy \over y} \left( 1 - \cos yt \right) \; p(y)
&\buildrel{t \to 
\infty}\over=& p(0) \left[ \ln(\mu \, t) + \gamma \right] + 
\int_0^{\infty} { dy \over y} \left[  p(y) -  p(0) \; \theta(\mu - y )
\right] + {\cal O}\left( {1 \over t } \right)  \; ,
\cr \cr  
\int_0^{\infty} { dy \over y} \left( t - {\sin yt \over y }\right) \;
p(y) &\buildrel{t \to 
\infty}\over=& t \; p(0) \left[ \ln(\mu \, t) + \gamma - 1 \right] + 
t \int_0^{\infty} { dy \over y} \left[  p(y) -  p(0) \; \theta(\mu - y )
\right] + {\cal O}\left( {1 \over t }   \right) \; , \nonumber
\end{eqnarray}

where $\gamma = 0.5772157 \ldots$ is  Euler's constant.

Notice that the formulas are {\bf independent} of the scale $ \mu $ as
one can easily see since  the derivative with respect to $ \mu $ of
the r. h. s. identically vanishes. The scale $ \mu $ has been
introduced just to have a dimensionless argument in the logs.

We have also used similar integrals when the resonance was away from
the threshold. We have for such case,
\begin{eqnarray}
\int_{-A}^{\infty} { dy \over y^2} \left( 1 - \cos yt \right) \; p(y)
&\buildrel{t \to \infty}\over=& \pi \; t \; p(0) + {\cal P}
\int_{-A}^{\infty} { dy \over y^2} \; [ p(y) -  p(0)] 
+  {\cal O}\left( {1 \over t } 
\right) \; ,  \cr \cr 
\int_{-A}^{\infty} { dy \over y} \left( t - {\sin yt \over y }\right) \;
p(y) &\buildrel{t \to  \infty}\over=& t \; {\cal P}
\int_{-A}^{\infty} { dy \over y} \; p(y) - \pi \, p'(0) +  {\cal
O}\left( {1 \over t }  \right)\; .\nonumber
\end{eqnarray}
where $ A $ is a fixed positive number. 

In addition, we needed in sec. III and IV integrals where the spectral
density has a logarithmic singularity at a finite point. 
\begin{eqnarray}\label{formulog}
\int_{-A}^{\infty} { dy \over y^2} \left( 1 - \cos yt \right) \;
p(y)\; \ln{|y| \over 2 {\bar T}}
&\buildrel{t \to \infty}\over=& \pi \; t \; p(0)  \left[ 1 - \gamma -
\ln(2\, t\,  {\bar T}) \right]  \cr \cr
&+& {\cal P}
\int_{-A}^{\infty} { dy \over y^2} \; [ p(y) -  p(0)] \; \ln{|y| \over
2 {\bar T}} +  {\cal O}\left( {1 \over t } 
\right) \; ,  \cr \cr 
\int_{-A}^{\infty} { dy \over y} \left( t - {\sin yt \over y }\right) \;
p(y) \; \ln{|y| \over 2 {\bar T}}
&\buildrel{t \to  \infty}\over=& t \; {\cal P}
\int_{-A}^{\infty} { dy \over y} \; p(y)\; \ln{|y| \over 2 {\bar T}} -
\pi \, p'(0)\left[\ln(2\, t\,  {\bar T}) + \gamma\right] +  {\cal 
O}\left( {1 \over t }  \right)\; .\nonumber
\end{eqnarray}


\section{\bf A simple example: the damped harmonic oscillator}
In this appendix we give a rather simple example of the dynamical renormalization group for pedagogical reasons 
and to illustrate the fundamental features within a simple setting. We consider the equation of motion of a damped
harmonic oscillator:
\begin{equation}
\ddot{y}+y=-\epsilon \dot{y}~, \epsilon <<1\label{damposc2}
\end{equation}
and seek a solution in a perturbative expansion in $\epsilon$ of the form $y=y_0+\epsilon y_1 + \epsilon^2 y_2 + \cdots$
where the $y_i$ are solutions to the following hierarchy of equations:
\begin{eqnarray}
\ddot{y}_0+y_0 &=& 0~,\nonumber\\
\ddot{y}_1+y_1 &=& -\dot{y}_0~,\nonumber\\
\ddot{y}_2+y_2 &=& -\dot{y}_1~,\nonumber\\
\vdots&\quad &\vdots\nonumber
\end{eqnarray}
These equations can be solved iteratively by starting from the zero
order solution 

$$
y_0(t)= A\; e^{it} + \mbox{c.c}\; ,
$$ 

in terms of the retarded Green's function

$$
G_{\text{ret}}(t-t')= \sin(t-t')\; \theta(t-t')~,
$$

Up to second order in $\epsilon$, the solution is given by
\begin{eqnarray}
y(t)&=&A\; e^{it}
\left[1-\frac{\epsilon}{2}
t+\frac{\epsilon^2}{8}\;t^2+i \; \frac{\epsilon^2}{8}t\right]+ \mbox{c.c.}
+\mbox{non-secular} \nonumber
\end{eqnarray}

Note that this solution contains secular terms that grow in $t$, the
terms denoted by {\em non-secular} remain finite at all times. We see
that the perturbative expansion breaks down at 
a time scale $\approx 1/\epsilon$. The expression in the brackets can
be interpreted as a change in the complex amplitude. 
The dynamical renormalization is achieved by introducing a time scale
$\tau$ at which the secular terms are absorbed in a 
renormalization of the complex amplitude. We write $ A=A(\tau)\;Z(\tau)
$ with $ Z(\tau)=1+\epsilon\; z_1(\tau) +\epsilon^2\; z_2(\tau) $ and choose
$z_i(\tau)$ to cancel the secular terms at the scale $\tau$, this is
similar to choosing the renormalization scale in the usual
renormalization program. Up to ${\cal O}(\epsilon^2)$ we find 

$$
z_1(\tau) = \frac{\tau}{2} \; ; \; z_2(\tau) = \frac{\tau^2}{8}-i \;
\frac{\tau}{8} \;.
$$

After renormalization the solution is given by 

\begin{eqnarray}
y(t,\tau)&=&A(\tau)\; e^{it}\left[1-\frac{\epsilon}{2}\;
(t-\tau) + \frac{\epsilon^2}{8}\;(t-\tau)^2
+i \; \frac{\epsilon^2}{8}\;(t-\tau)\right]+ \mbox{c.c.}+\mbox{nonsecular}\;.
\label{dampsolu} 
\end{eqnarray}

Since $\tau$ is an arbitrary scale, the solution cannot depend on it,
thus the statement 
$ dy(t,\tau)/d\tau =0 $  leads to the dynamical renormalization group
equation to this order 

$$
\frac{\partial A(\tau)}{\partial \tau}
+A(\tau)\left(\frac{\epsilon}{2}-i \; \frac{\epsilon^2}{8}\right)=0 
$$
where we have expanded the $\tau$ derivative of the amplitude in a power series expansion in $\epsilon$ consistently to
second order. Obviously the solution to the renormalization group equation is given by
$$
A(\tau)= A(0) \;  e^{-\frac{\epsilon}{2}\tau} \;  e^{i
\frac{\epsilon^2}{8}\tau} 
$$

setting $t=\tau$ in (\ref{dampsolu}) we finally find

$$
y(t) = A(0) \;  e^{-\frac{\epsilon}{2}\tau} \;  e^{i
(1-\frac{\epsilon^2}{8})t}+\mbox{c.c.} 
$$

which is obviously the correct solution to second order. Further simple and not-so-simple examples can be found in ref.\cite{goldenfeld}.



\begin{figure}[t] 
\epsfig{file=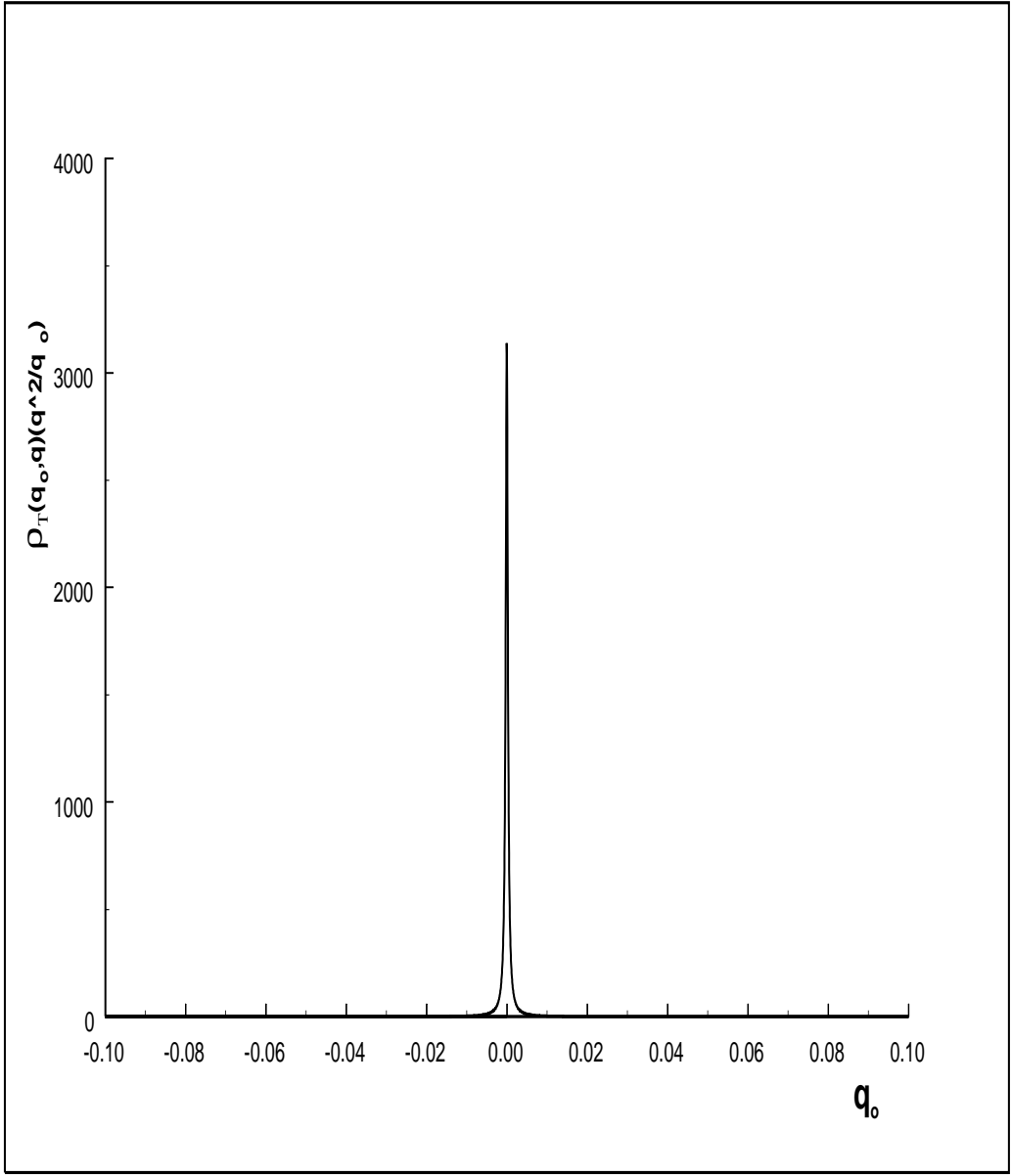,width=15cm,height=18cm} 
\caption{ $\rho_T(q_0,q)(q^2/q_0)$ for $q=0.1 \frac{eT}{\sqrt{12}}$ vs. $q_0$ in units of $\frac{eT}{\sqrt{12}}$   \label{fig2}}
\end{figure} 
\end{document}